\catcode`\@=11
\newif\if@fewtab\@fewtabtrue

{\count255=\time\divide\count255 by 60
\xdef\hourmin{\number\count255}
\multiply\count255 by-60\advance\count255 by\time
\xdef\hourmin{\hourmin:\ifnum\count255<10 0\fi\the\count255}}
\def\ps@draft{\let\@mkboth\@gobbletwo
    \def\@oddhead{}
    \def\@oddfoot
       {\hbox to 7 cm{$\scriptstyle Draft\ version:\ \draftdate$
       \hfil}\hskip -7cm\hfil\rm\thepage \hfil}
    \def\@evenhead{}\let\@evenfoot\@oddfoot}

\def\ceqno{\global\@fewtabfalse
    \ifcase\@eqcnt \def\@tempa{& & &}\or \def\@tempa{& &}
      \or \def\@tempa{&}
      \or\def\@tempa{}\fi\@tempa
{\rm(\theequation)}}

\def\aeqno#1{\global\@fewtabfalse
    \ifcase\@eqcnt \def\@tempa{& & &}\or \def\@tempa{& &}
      \or \def\@tempa{&}
      \or\def\@tempa{}\fi\@tempa
{\rm(\theequation,#1)}}

\def\label#1{\ifnum\draftcontrol=1
 \global\def\draftnote{$\scriptstyle #1$}\fi
 \@bsphack\if@filesw {\let\thepage\relax
   \def\protect{\noexpand\noexpand\noexpand}%
\xdef\@gtempa{\write\@auxout{\string
      \newlabel{#1}{{\@currentlabel}{\thepage}}}}}\@gtempa
   \if@nobreak \ifvmode\nobreak\fi\fi\fi
  \@esphack}

\def\alabel#1#2{\label{#1}\global\@fewtabfalse
    \ifcase\@eqcnt \def\@tempa{& & &}\or \def\@tempa{& &}
      \or \def\@tempa{&}
      \or\def\@tempa{}\fi\@tempa
{\hbox to 3cm{\phantom{\rm(\theequation,#2)}
\draftnote \hfil}\hskip -3cm {\rm(\theequation,#2)}}}

\def\clabel#1{\label{#1}\global\@fewtabfalse
    \ifcase\@eqcnt \def\@tempa{& & &}\or \def\@tempa{& &}
      \or \def\@tempa{&}
      \or\def\@tempa{}\fi\@tempa
{\hbox to 3cm{\phantom{\rm(\theequation)}
\draftnote \hfil}\hskip -3cm{\rm(\theequation)}}}

\def\eqnarray{\def\draftnote{{}}\global\@fewtabtrue
\stepcounter{equation}\let\@currentlabel=\theequation
\global\@eqnswtrue
\global\@eqcnt\z@\tabskip\@centering\let\\=\@eqncr
$$\halign to \displaywidth\bgroup\@eqnsel\hskip\@centering\@eqcnt\z@
  $\displaystyle\tabskip\z@{##}$&\global\@eqcnt\@ne
  \hskip 1\arraycolsep \hfil${##}$\hfil
  &\global\@eqcnt\tw@ \hskip 1\arraycolsep
$\displaystyle\tabskip\z@{##}$
\hfil  \tabskip\@centering&\global\@eqcnt\thr@@\llap{##}\tabskip\z@
\cr}

\def\endeqnarray{\@@eqncr\egroup
      \global\advance\c@equation\m@ne$$\global\@ignoretrue}

\def\@eqnnum{\hbox to 3cm{\phantom{\rm(\theequation)} \draftnote
                         \hfil}\hskip -3cm {\rm(\theequation)}}

\def\@@eqncr{\let\@tempa\relax
    \ifcase\@eqcnt \def\@tempa{& & &}\or \def\@tempa{& &}
      \or \def\@tempa{&}
      \or\def\@tempa{}
\fi\@tempa
\if@eqnsw
\if@fewtab\@eqnnum\fi
\stepcounter{equation}\fi\global
\@eqnswtrue\global\@eqcnt\z@\global\@fewtabtrue\cr}

\def\draftcite#1{\ifnum\draftcontrol=1#1\else{}\fi}

\def\@lbibitem[#1]#2{\item{}\hskip -3cm \hbox to 2cm
{\hfil$\scriptstyle\draftcite{#2}$}\hskip
1cm[\@biblabel{#1}]\if@filesw
     {\def\protect##1{\string ##1\space}\immediate
      \write\@auxout{\string\bibcite{#2}{#1}}}\fi\ignorespaces}

\def\@bibitem#1{\item\hskip -3cm \hbox to 2cm
{\hfil $\scriptstyle\draftcite{#1}$}\hskip 1cm
\if@filesw \immediate\write\@auxout
       {\string\bibcite{#1}{\the\value{\@listctr}}}\fi\ignorespaces}

\def\nsection#1{\section{#1}\setcounter{equation}{0}}

\def\nappendix#1{\vskip 1cm\no{\bf Appendix #1}\def\thesection{#1}
\setcounter{equation}{0}}

\font\tendl=msbm10  scaled \magstep1%double line
\font\sevendl=msbm7 scaled \magstep1
\font\fivedl=msbm5 scaled \magstep1
\font\tengl=eufm10  scaled \magstep1% gothic letters
\font\sevengl=eufm7 scaled \magstep1
\font\fivegl=eufm5 scaled \magstep1

\newfam\dlfam  % \dl is double line
\textfont\dlfam=\tendl \scriptfont\dlfam=\sevendl
\scriptscriptfont\dlfam=\fivedl
\newfam\glfam  % \gl is gothic letters
\textfont\glfam=\tengl \scriptfont\glfam=\sevengl
\scriptscriptfont\glfam=\fivegl

\def\draftdate{\number\month/\number\day/\number\year\ \ \ \hourmin }

\global\def\draftcontrol{0}
\catcode`\@=12
\def\tilde{\widetilde}
\def\hat{\widehat}
\documentstyle[12pt]{article}

\def\theequation{{\thesection.\arabic{equation}}}

\newcommand{\be}{\begin{eqnarray}}
\newcommand{\en}{\end{eqnarray}\vs 0.5 cm}

\newcommand{\no}{\noindent}
\newcommand{\vs}{\vskip}
\newcommand{\hs}{\hspace}

\newcommand{\p}{\partial}

\newcommand{\NR}{{{\bf R}}}%letra doble raya en modo matematico
\newcommand{\NC}{{{\bf C}}}%letra doble raya en modo matematico
\newcommand{\Nx}{{{\bf x}}}
\newcommand{\Ny}{{{\bf y}}}

\newcommand{\Nu}{{{\bf u}}}
\newcommand{\Ns}{{{\bf s}}}
\newcommand{\Nt}{{{\bf t}}}
\newcommand{\Nz}{{{\bf z}}}

\newcommand{\Np}{{{\bf p}}}

\newcommand{\qq}{\begin{eqnarray}}

\newcommand{\da}{\partial}
\newcommand{\ee}{{\rm e}}

\newcommand{\qqq}{\end{eqnarray}}

\newcommand{\la}{\lambda}

\newcommand{\CC}{{\cal C}}
\newcommand{\CD}{{\cal D}}

\newcommand{\CF}{{\cal F}}
\newcommand{\CG}{{\cal G}}
\newcommand{\CH}{{\cal H}}

\newcommand{\CJ}{{\cal J}}
\newcommand{\CK}{{\cal K}}
\newcommand{\CL}{{\cal L}}
\newcommand{\CM}{{\cal M}}

\newcommand{\CO}{{\cal O}}
\newcommand{\CP}{{\cal P}}

\newcommand{\CR}{{\cal R}}
\newcommand{\CS}{{\cal S}}
\newcommand{\CT}{{\cal T}}
\newcommand{\CU}{{\cal U}}
\newcommand{\CV}{{\cal V}}

\newcommand{\s}{\hspace{0.05cm}}
\newcommand{\m}{\hspace{0.025cm}}

\newcommand{\hf}{{_1\over^2}}

\begin{document}
\title{\bf{Slow modes \s in \s passive advection}}
\author{\\
Denis Bernard \\C.N.R.S., Service de Physique Theorique
de CEA,\\91191 Gif-sur-Yvette, France\\
\\
Krzysztof Gaw\c{e}dzki \\C.N.R.S., I.H.E.S., 
91440 Bures-sur-Yvette, France\\ 
\\
Antti Kupiainen
\\Mathematics Department, Helsinki University, 
\\PO Box 4, 00014 Helsinki, Finland
\\}

\date{ }
\maketitle
\begin{abstract}
\vskip 0.2cm
\noindent The anomalous scaling in the Kraichnan model
of advection of the passive scalar by a random 
velocity field with non-smooth spatial behavior is 
traced down to the presence of slow resonance-type
collective modes of the stochastic evolution of fluid 
trajectories. We show that the slow modes are organized into
infinite multiplets of descendants of the primary 
conserved modes. Their presence is linked to the non-deterministic 
behavior of the Lagrangian trajectories at high Reynolds numbers
caused by the sensitive dependence on initial conditions 
within the viscous range where the velocity fields are more regular. 
Revisiting the Kraichnan model with smooth velocities we describe 
the explicit solution for the stationary state of the scalar. 
The properties of the probability distribution function 
of the smeared scalar in this state are related 
to a quantum mechanical problem involving the Calogero-Sutherland 
Hamiltonian with a potential.
\hfill
\\
\end{abstract}

%%%%%%%%%%%%%%%%%%%%%%%%%%%%%%%%%%%%%%%%%%%%%%%%%%%%%%%%%%%%%%%%
%%% for draft versions, suppress in definitive version
%\draft
%%
%%%%%%%%%%%%%%%%%%%%%%%%%%%%%%%%%%%%%%%%%%%%%%%%%%%%%%%%%%%%%%%%
%\begin{center}
%(draft, May 9, 1997)
%\end{center}
%\vs 0.6cm
\nsection{Introduction}

One of the basic open problems in fully developed
hydrodynamical turbulence is the understanding
of the origin of observed violations of the Kolmogorov
\cite{Kolm} scaling. The violations indicate presence 
of strong short-distance intermittency in the turbulent 
cascade, i.e.\s\s of frequent occurrence of large fluctuations 
on short distances. Recently some progress has been
achieved in the understanding of the analogous problem
for the passive advection of a scalar quantity  
by a random velocity field. The scalar is known 
to exhibit strong short-distance intermittency 
even if such is absent in the velocity field.
In the simplest model of the passive scalar,
due to Kraichnan \cite{Kr68}, one assumes a Gaussian 
distribution of time-decorrelated and spatially 
non-smooth velocities. The anomalous scaling 
of the scalar in this model was related in references 
\cite{SS}\cite{Falk}\cite{GK} to zero modes of differential 
operators describing the stochastic evolution of the flow. 
In the present paper we elaborate on this idea showing that 
the short-distance intermittency of the scalar is due
to the presence of slow collective modes in the otherwise 
super-diffusive evolution of the (quasi)-Lagrangian 
trajectories of fluid particles. We show that 
in the Kraichnan model the slow modes, reminiscent 
of resonances in multi-body problems, are organized 
into infinite multiplets of descendants with the zero 
modes playing the role of primary objects. This structure 
might indicate the presence of hidden infinite symmetries 
in the Kraichnan problem.
\vskip 0.3cm

The other important feature of the Lagrangian flow in non-smooth 
velocities is its intrinsically probabilistic character: 
the Lagrangian trajectories of the fluid particles
behave randomly even in a fixed velocity field. This phenomenon 
appears to be closely related to the presence of the 
slow modes in the stochastic flow of fluid particles.
In more realistic velocity fields which are regularized 
on the viscous scale the effective stochasticity of the fluid 
trajectories is due to their sensitive dependence on initial 
conditions on scales shorter than the viscous one. 
We expect both phenomena: the presence of resonant slow modes
in the Lagrangian flow and the non-deterministic character
of the fluid trajectories, to be present in more realistic
high Reynolds number velocity ensembles and to be responsible 
for their intermittency.
\vskip 0.3cm

The version of the Kraichnan model with smooth velocity fields, 
relevant for the description of the distances smaller than 
the viscous scale, has been intensively studied, 
see \cite{SS} and \cite{SS1} to \cite{FKLM}. 
We return to this case developing further the tools
of harmonic analysis used first for this model in \cite{SS} 
and \cite{SS2}. These tools allow a fast calculation
of the the Lyapunov exponents for the flow of fluid
particles found first in \cite{CFKL} and \cite{CGK}.
We also explicitly construct the stationary state 
of the scalar relating its functional Fourier transform 
to a certain Schr\"{o}dinger operator on the symmetric 
space \s$SL(d)/SO(d)\s$ where \s$d\s$ is the space dimension.
In particular, we compute the exponential decay 
rate of the probability distribution functions \s$p(\theta)\s$ 
of smeared scalar values obtaining in three dimensions (and above)
a result different from that of \cite{CFKL}\cite{CGK}\cite{FKLM}. 
The discrepancy is traced to the contribution 
of correlations of different (pairs of) fluid trajectories 
disregarded in \cite{CFKL}\cite{CGK}\cite{FKLM}.
For the exponential decay rate of the Fourier transform 
of \s$p(\theta)\s$ our results reproduce fully the calculations
of \cite{CFKL}\cite{CGK} and confirm their semiclassical
interpretation proposed in \cite{FKLM}.
\vskip 0.5cm

The paper is organized as follows. In Sect.\s\s2
we present the Kraichnan model and obtain its solution
employing a path integral formalism. Sect.\s\s3 recalls
briefly the analysis of \cite{GK} and \cite{BGK}
establishing anomalous scaling of the scalar
by perturbative analysis of the scaling zero modes 
of operators governing the flow. 
The physical interpretation of the zero modes 
as scaling structures conserved in mean is the subject 
of Sect.\s\s4. Sect.\s\s5 discusses the collective 
slow modes of the random flow of fluid particles.
The analytic origin of the slow modes is unraveled in 
more technical Sect.\s\s6. Sect.\s\s7 describes the intricacies 
of the probabilistic description of fluid trajectories. 
{}Finally, Sects.\s\s8 and 9 study in detail the case of Kraichnan
model with smooth velocity field elaborating on the earlier
results of \cite{SS} and of \cite{SS1} to \cite{FKLM}. 
Appendix A explicitly analyzes the slow modes in the relative 
motion of two fluid particles in non-smooth velocity field. 
Appendix B contains some more details on the smooth velocity 
case related to the results of \cite{CFKL} and \cite{BCKL}.
\vskip 0.8cm

\nsection{Kraichnan model of passive scalar}

Let us consider an advection of a scalar quantity 
\s$T(t,x)\s$ (the temperature) in $d$ space dimensions.  
The time evolution of \s$T\s$ is governed by the linear  equation
\qq
\da_t T\s+\s v\cdot\nabla\s T\s-\s\kappa\s\Delta T\s=\s f
\label{jeden}
\qqq
where \s$v(t,x)\s$ is the incompressible (\m$\nabla\cdot v=0\m$)
velocity field of the advecting fluid, \s$\kappa\s$ is the diffusion 
constant and \s$f(t,x)\s$ is a given source term. Denote by 
\s$R(t,x;t_0,x_0)\s$ the solution of the homogeneous equation
\qq
(\da_t\m+\m v\cdot\nabla\m-\m\kappa\m\Delta)\s\m R(t,t_0)\s=\s0
\qqq
with the initial condition
\qq
R(t_0,x;t_0,x_0)\s=\s\delta(x-x_0)\s.
\qqq
We shall call \s$R(t,x;t_0,x_0)\s$ the evolution kernel
and the corresponding operator \s$R(t,t_0)\s$ the evolution 
operator. The solution of Eq.\s\s(\ref{jeden}) has the form
%\qq
%T(t)\s=\s R(t,t_0)\s\m T(t_0)
%\s+\s\int_{_{t_0}}^{^t} R(t,s)\s f(s)\s\m ds
%\qqq
%which is the shorthand version of
\qq
T(t,x)\s=\s\int R(t,x;t_0,y)\s T(t_0,y)\s\m dy
\s+\s\int_{_{t_0}}^{^t}\int R(t,x;s,y)\s f(s,y)\s\m ds\s dy
\label{solu}
\qqq
with \s$T(t_0)\s$ being the initial configuration of \s$T\m$
at time \s$t_0\m$.
\vskip 0.5cm

There exists a functional integral formula for the evolution 
kernel which, for sufficiently regular \s$v\m$, \m  may 
be easily given a rigorous sense as an integral with respect 
to the Wiener measure with density:
\qq
R(t,x;t_0,x_0)\s=\s\int\limits_{x(t_0)=x_0\atop x(t)=x}
\ee^{-{1\over 4\kappa}\int_{t_0}^tds\s[\m^{\cdot}\hs{-0.14cm}x(s)-
v(s,x(s))]^2}\s\s Dx
\label{position}
\qqq
where \s${}^{\cdot}\hs{-0.16cm}x\equiv{{dx}\over{dt}}\m$. 
\m It will be useful to rewrite the above 
functional integral as a phase space one:
\qq
R(t,x;t_0,x_0)\s=\s\int\limits_{x(t_0)=x_0\atop x(t)=x}
\ee^{-\int_{t_0}^tds\s[\kappa\m p(s)^2\s+\s i\s p(s)\cdot(
\m^{\cdot}\hs{-0.14cm}x(s)-v(s,x(s)))]}\s\s Dx\s\m Dp
\label{phase}
\qqq
with the Gaussian integral over the unconstrained paths 
\s$s\mapsto p(s)\s$ reproducing the previous integral.
{}From the functional integral representations it is clear that 
when \s$\kappa\to0\s$ then 
\qq
R(t,x;t_0,x_0)\ \s\rightarrow\s\ \delta(x-x(t;t_0,x_0))
\label{ltr}
\qqq
where \s$x(t;t_0,x_0)\s$ is the Lagrangian trajectory
of the fluid particle satisfying the equations
\qq
{}^{\cdot}\hs{-0.18cm}x\s=\s v(t,x)\s,\quad\ \ \ x(t_0)\s=\s x_0\s.
\label{lagr}
\qqq
Indeed, we may set \s$\kappa=0\s$ in the phase space integral
(\ref{phase}) and the \s$p$-integral gives then
a delta function(al) concentrated on the Lagrangian trajectory.
\s For small positive \s$\kappa\m$, \m on the other hand, 
\s$R(t,x;t_0,x_0)\s$ is the probability distribution function 
(p.d.f.) of the endpoint of a small 
Brownian motion around the Lagrangian trajectory. 
Such a Brownian motion \s$x_\beta(t;t_0,x_0)\s$ with a drift 
is a solution of the stochastic 
ordinary differential equation (ODE) 
\qq
dx\s=\s v(t,x)\m dt\s+\s\kappa\m d\beta\s,\quad\ \ \ 
x(t_0)=x_0 
\label{stoch}
\qqq
with \s$\beta(t)\s$ denoting the Brownian motion
without drift. Thus
\qq
R(t,x;t_0,x_0)\s=\s E(\s\delta(x-x_\beta(t;t_0,x_0))\m) 
\label{expect}
\qqq
where \s$E(\s\m\cdot\s\m)\s$ denotes the expectation with respect
to the Wiener measure of \s$\beta\m$. \m
Eq.\s\s(\ref{expect}) is another form of Eq.\s\s(\ref{position}).
\vskip 0.5cm

We shall be interested in the situation when both
velocities \s$v\s$ and source \s$f\s$ in Eq.\s\s(\ref{jeden})
are random so that Eq.\s\s(\ref{jeden}) is a stochastic
PDE. In order to solve such a stochastic equation, we should define the 
evolution kernel \s$R(t,x;t_0,x_0)\s$ as a random, \s$v$-dependent 
process or, in plain English, be able to compute various expectation 
values of \s$R\m$'s like
\qq
\CP_n(\Nt,\Nx;\Nt_0,\Nx_0)\ \equiv\ \langle\s\prod\limits_{i=0}^n
R(t_i,x_i;t_{0,i},x_{0,i})\s\rangle\ =\ 
E(\s\prod\limits_{i=0}^n\delta(x_i-x_{\beta_i}(t_i;t_{0,i},x_{0,i})\m)
\label{jpd}
\qqq
where \s$\Nt\equiv(t_1,\dots,t_n)\s$, \s$\Nx\equiv(x_1,\dots,x_n)\s$ 
and similarly for \s$\Nt_0\m$, \s$\Nx_0\m$ and where 
\s$E(\s\m\cdot\s\m)\s$ denotes the expectation 
w.r.t. $\beta_i\m$'s and \s$v\m$. 
\m It is clear from the second expression for \s$\CP_n(\Nt,\Nx;\Nt_0,
\Nx_0)\s$ that it gives the joint p.d.f. of the ends \s$x_{\beta_i}
(t_i;t_{0,i},x_{0,i})\s$ 
of \s$n\s$ Brownian motions (independent for given 
\s$v\m$) \m around the Lagrangian trajectories starting at times 
\s$t_{0,i}\s$ from points \s$x_{0,i}\m$. \s The \s$\kappa\to0\s$ 
limit of \s$\CP_n(\Nt,\Nx;\Nt_0,\Nx_0)\s$ (if exists) should simply 
give the joint p.d.f. of the endpoints \s$x(t_i;t_{0,i},x_{0,i})\s$ 
of \s$n\s$ Lagrangian trajectories\footnote{K.G. thanks Ya. Sinai
for attracting his attention to the statistics of Lagrangian
trajectories}.
\vskip 0.5cm

Let us assume that the velocity is a Gaussian
stationary field with mean zero and covariance
\qq
\langle\m v^\alpha(t_1,x_1)\s v^\beta(t_2,x_2)\m\rangle
\s=\s D^{\alpha\beta}(t_{12},\m x_{12})
\qqq
where \s$t_{12}\equiv t_1-t_2\m,\ \m x_{12}\equiv x_1-x_2\s$ and 
\s$\da_\alpha D^{\alpha\beta}(t,x)=0\s$ to assure the
incompressibility. Employing the phase 
space path integral representation (\ref{phase}) 
and performing the Gaussian functional integration over \s$v\m$,
\m we obtain\footnote{similar expressions appeared in 
\cite{Chert}\cite{FKL}}:
\qq
&&\CP_n(\Nt,\Nx;\Nt_0,\Nx_0)\ =\ \langle
\hs{-0.3cm}\int\limits_{x_i(t_{0,i})=x_{0,i}
\atop x_i(t_i)=x_i}\hs{-0.6cm}
\ee^{-\sum_i\{\int_{t_{0i}}^{t_i}ds_i\s[\m\kappa\m 
p_i(s_i)^2\s+\s i\s p_i(s_i)\cdot(\m^{\cdot}\hs{-0.14cm}x_i(s_i)
-v(s_i,x_i(s_i))\m)\m]\s\}}
\s\s D\Nx\s\m D\Np\ \s\rangle\cr\cr\cr
&&\hs{0.5cm}=\hs{-0.3cm}\int\limits_{x_i(t_{0,i})=x_{0,i}
\atop x_i(t_i)=x_i}\hs{-0.6cm}\ee^{-\sum_i
\{\int_{t_{0i}}^{t_i}ds_i\s[\m\kappa\m 
p_i(s_i)^2\s+\s i\s p_i(s_i)\cdot\s^{\cdot}\hs{-0.14cm}x_i
(s_i)\m]\s\}}\cr
&&\hs{2.5cm}\cdot\ \ee^{-\s{1\over 2}\sum_{i,j}\int_{t_{0i}}^{t_i}
ds_i\int_{t_{0j}}^{t_j}ds_j
\s\s D^{\alpha\beta}(s_i-s_j,\s x_i(s_i)-x_j(s_j))\s\m 
p_i^\alpha(s_i)\m p_j^\beta(s_j)}\s\s D\Nx\s\m D\Np\s. 
\label{expp}
\qqq
If, following Kraichnan \cite{Kr68}, we assume
that \s$v(t,x)\s$ is also decorrelated in time, i.e. that
\qq
D^{\alpha\beta}(t,x)\s=\s\delta(t)\s\m\CD^{\alpha\beta}(x)\s,
\label{cov}
\qqq
then formula (\ref{expp}) for \s$\CP_n(\Nt,\Nx;\Nt_0,\Nx_0)\s$
may be further simplified. Let as set all \s$t_i\m$ equal to \s$t\s$
and all \s$t_{0i}\m$ equal to $t_0\m$. \m We shall 
denote the corresponding \s$\CP_n\s$
by \s$\CP_n(t,\Nx;t_0,\Nx_0)\s$ (the general case can be reconstructed
from the special one for \s$v\m$'s delta-correlated in time).
The fundamental property of the p.d.f.'s \s$\CP_n(t,\Nx;t_0,\Nx_0)\s$ 
for the time-decorrelated velocities (not necessarily Gaussian)
is the composition property
\qq
\int \CP_n(t,\Nx;s,\Ny)\s\m \CP_n(s,\Ny;t_0,\Nx_0)\s\s d\Ny\ =\ 
\CP_n(t,\Nx;t_0,\Nx_0)\s.
\label{sgp}
\qqq
{}From expression (\ref{expp}) we obtain, assuming relation (\ref{cov}),
\qq
&&\hspace{1cm}\CP_n(t,\Nx;t_0,\Nx_0)\ \cr\cr
&&=\hs{-0.5cm}\int\limits_{x_i(t_{0,i})=x_{0,i}
\atop x_i(t_i)=x_i}\hs{-0.6cm}\ee^{-\int_{t_{0}}^{t}ds\s\m
[\m\kappa\m\sum_ip_i(s)^2\s+\s{1\over 2}\sum_{i,j}
\CD^{\alpha\beta}(\m x_i(s)-x_j(s))\s\m 
p_i^\alpha(s)\m p_j^\beta(s)\s+\s
i\sum_i p_i(s)\cdot\s^{\cdot}\hs{-0.14cm}x_i
(s)\m]\s}\s\s D\Nx\s\m D\Np\s.\hs{1.3cm}
\label{expp1}
\qqq
It is easy to see that the right hand side is a 
phase space path integral
expression for the heat kernel (dynamical Green function) 
of the 2$^{\rm\m nd}$  order 
(positive, elliptic) differential operator
\qq
\CM_n\ =\ -\m{_1\over^2}\sum\limits_{i,j=1}^n
\CD^{\alpha\beta}(x_{ij})\s\m\da_{x_i^\alpha}\da_{x_j^\beta}
\s-\s\kappa\sum\limits_{i=1}^n\Delta_{x_i}\s,
\label{Mn}
\qqq
i.e. that
\qq
\CP_n(t,\Nx;t_0,\Nx_0)\ =\ \ee^{-(t-t_0)\m \CM_n}(\Nx,\Nx_0)\s,
\label{hker}
\qqq
compare Eq.\s\s(\ref{phase}). Note that, due to incompressibility,
there is no ordering ambiguity in passing from the path integral
to the expression for \s$\CM_n\m$. Rigorously minded person may 
take expressions (\ref{hker}) as defining the evolution
operators for the stochastic PDE equation (\ref{jeden})
in the time decorrelated case. Of course, the composition property
(\ref{sgp}) follows from the semigroup law for the heat kernels.
\vskip 0.5cm

Let us now go back to the passive scalar.
Assume that both \s$v\s$ and \s$f\s$ are independent
stationary processes. Imposing also the zero initial
condition for \s$T\s$ at \s$t_0=-\infty\m$,
we obtain using Eq.\s\s(\ref{solu}) the following expression
for the correlators of \s$T\m$:
\qq
\langle\s\prod\limits_{i=0}^n T(t_i,x_i)\m\rangle\s=\s
\prod\limits_{i=1}^n\int_{_{-\infty}}^{^{t_i}}\hs{-0.15cm}
ds_i\int dy_i
\s\s\s \CP_n(\Nt,\Nx;\Ns,\Ny)\s\s\langle\m\prod\limits_{i=1}^n
f(s_i,y_i)\m\rangle\s.
\label{scf}
\qqq
It should be clear that if a stationary state of the scalar
is generated for large time and independent of the initial
conditions (say, decaying at infinity) then its correlation
functions should be given by the above equation. Hence
the importance of understanding the behavior of the p.d.f.'s
\s$\CP_n(\Nt,\Nx;\Nt_0,\Nx_0)\m$.
\vskip 0.5cm

Assume now that the source \s$f\s$ (independent of \s$v\m$) \m is 
a Gaussian process with mean zero and covariance
\qq
\langle\m f(t_1,x_1)\s f(t_2,x_2)\m\rangle\s
=\s\delta(t_{12})\s\m \CC(x_{12})
\qqq
where \s$\CC\s$ is a positive definite test function. 
In this case and for the Gaussian, time decorrelated
velocities, Eqs.\s\s(\ref{scf}) simplify permitting
an inductive calculation of the stationary correlation
functions of the scalar. Let us see how this works for
equal time correlators. We may consider only
the even-point functions of \s$T\s$, \s$\CF_{2n}(\Nx)\s\equiv
\s\langle\m T(t,x_1)\s\cdots\s T(t,x_{2n})\m\rangle\m$, \m 
since the odd correlators
of \s$f\s$ vanish implying the same property of the 
\s$T\s$ correlators. For the 2-point function we obtain
\qq
\CF_2(x_{12})\ =\ \int_{_{-\infty}}^{^{t}}\hs{-0.15cm}ds\int d\Ny\s\s
\s\ee^{-(t-s)\m \CM_2}(\Nx,\Ny)
\s\s\CC(y_{12})\ =\ \int d\Ny\s\s\s \CM_2^{\m-1}(\Nx,\Ny)\s\m\CC(y_{12})
\qqq
and for the 4-point function
\qq
\CF_4(\Nx)\ =\ \sum_{1\leq i<j\leq 4}\hs{-0.05cm}
\int_{_{-\infty}}^{^{t}}\hs{-0.15cm}ds
\int_{_{-\infty}}^{^{s}}\hs{-0.15cm}ds'\int d\Ny
\int d\Nz\ \s\ee^{-(t-s)\m \CM_4}(\Nx,\Ny)
\s\s\CC(y_{ij})\s\s\cr
\cdot\s\s\ee^{-(s-s')\m \CM_2}(y_1,\mathop{\dots}
\limits_{\hat i\s\s\hat j},y_4,\Nz)\s\s\CC(z_{12})\s\s\cr
=\ \sum_{1\leq i<j\leq 4}\hs{-0.05cm}\int\CM_4^{\m-1}
(\Nx,\Ny)\s\s \CF_2(y_1,\mathop{\dots}\limits_{\hat i
\s\s\hat j},y_4)\s\s\CC(y_{ij})\ d\Ny\s.
\label{4pf}
\qqq
Similar arguments for the $2n$-point function give
\qq
\CF_{2n}(\Nx)\  
=\ \sum_{1\leq i<j\leq 2n}\hs{-0.05cm}\int\CM_{2n}^{\m-1}
(\Nx,\Ny)\s\s \CF_{2n-2}(y_1,\mathop{\dots}\limits_{\hat i
\s\s\hat j},y_{2n})\s\s\CC(y_{ij})\ d\Ny\s.
\label{2npf}
\qqq
The above equations permit an inductive calculation
of the stationary equal time correlation functions
of \s$T\m$ with the use of the (static) Green functions 
\s$\CM_n^{\m-1}(\Nx,\Ny)\s$ of operators \s$\CM_n\m$.
\vskip 0.8cm

\nsection{Zero mode dominance}

We shall be interested in the case 
where the spatial part \s$\CD^{\alpha\beta}(x)\s$ of the 
\s$v$-covariance has the form
\qq
\CD^{\alpha\beta}(x)\s
=\s \CD_0\delta^{\alpha\beta}\s-\s d^{\alpha\beta}(x)
\label{cov1}
\qqq
where \s$d^{\alpha\beta}(x)\s$ scales with power \s$2-\gamma\m$, 
\qq
d^{\alpha\beta}(x)\ \sim\ \vert x\vert^{2-\gamma}\s,
\label{asy}
\qqq
for small \s$\vert x\vert\m$. Here \s$0\leq\gamma\leq 2\s$
is a fixed parameter. The tensorial form of \s$d^{\alpha\beta}(x)\s$
is fixed for small \s$\vert x\vert\s$ by the incompressibility 
condition \s$\da_\alpha d^{\alpha\beta}(x)=0\m$:
\qq
d^{\alpha\beta}(x)\ \cong\ {_D\over^{d-1}}
\s(\m(d+1-\gamma)\m\delta^{\alpha\beta}
\m\vert x\vert^{2-\gamma}\s-\s({2-\gamma})\s 
x^\alpha x^\beta\m\vert x\vert^{-\gamma}\m)
\s\equiv\s d^{\alpha\beta}_{\rm sc}(x) 
\label{asy1}
\qqq
where \s$D\s$ is a constant. For \s$0<\gamma<2\m$, \m one may take
\qq
\CD^{\alpha\beta}(x) \ \sim\ \int{{\ee^{-i\m k\cdot x}}\over
{(k^2+m^2)^{(d+2-\gamma)/2}}}\s\m(\delta^{\alpha\beta}-{{k^\alpha
k^\beta}\over{k^2}})\s\m dk
\label{cv2}
\qqq
where \s$m\s$ is an infrared regulator. Relations 
(\ref{asy}) and (\ref{asy1}) hold then for
\s$m\vert x\vert\ll 1\m$. \m When \m$m\to0\m$, 
\s$d^{\alpha\beta}(x)\s$ tends to the scaling form
\s$d_{\rm sc}^{\alpha\beta}(x)\s$ but 
$\CD_0$ diverges like \m$\CO(m^{\gamma-2})\m$.
\vskip 0.5cm

The Gaussian distribution with covariance given 
by Eqs.\s\s(\ref{cov}) and (\ref{cov1}) is 
relatively far from a realistic description of the statistics 
of turbulent flows. First, it excludes the velocity intermittency, 
i.e.\s\s more frequent occurrence of large deviations of velocity 
differences than in the normal distribution. Such occurrence 
characterizes short scales in the inertial interval 
of the turbulent cascade. 
Second, the time decorrelation 
is a brutal approximation since one observes scale-dependent
time correlations in turbulent flows. The power-law growth
of the velocity difference covariance
\qq
\langle\m(v^\alpha(t_1,x_1)-v^\alpha(t_1,x_2))\s(v^\beta(t_2,x_1)
-v^\beta(t_2,x_2))
\m\rangle&=&2\s\delta(t_{12})\s\m d^{\alpha\beta}(x_{12})\cr\cr
&\sim&\s\delta(t_{12})\s\m\vert x_{12}\vert^{2-\gamma}
\qqq
mimics, however, the expected behavior in the turbulent cascade
(the Kolmogorov value of the scaling exponent corresponds to
\s$\gamma={2\over 3}\s$ since time appears to scale like length 
to power \s$\gamma\s$ in the model). The point is that even 
the velocity distributions far from realistic, as the one described 
above, induce strongly intermittent scalar distributions and 
the purpose of the Kraichnan model is to understand this 
phenomenon in the simplest context.
\vskip 0.5cm

Operators \s$\CM_n\s$ may be rewritten in the form
\qq
\CM_n\ =\ \m\sum\limits_{1\leq i<j\leq n}\hs{-0.1cm}
d^{\alpha\beta}(x_{ij})\s\m\da_{x_i^\alpha}\da_{x_j^\beta}
\s-\s\kappa\sum\limits_{i=1}^n\Delta_{x_i}
\s-\s{_1\over^2}\s \CD_0\m(\sum\limits_{i=1}^n\da_{x_i^\alpha})^2
\label{Mn1}
\qqq
where the last term drops out in the action on translationally 
invariant functions. We shall, somewhat pedantically, denote 
the operator \s$\CM_n\s$ acting in the translational invariant 
sector by \s$M_n\m$. \m We shall view \s$M_n\s$ 
as an operator in the reduced space \s$L^2(\NR^{d_n})\m$,
with \m$d_n\equiv (n-1)d\m$. \m This is the space 
of functions of the difference variables 
\s$x_{in}\equiv x_i-x_n\m$, \m square-integrable with  
the measure \s$d'\Nx\equiv dx_{1n}\cdots dx_{(n-1)\m n}\m$.
The heat kernel of \s$M_n\m$,
\qq
\ee^{-(t-t_0)\m M_n}(\Nx,\Nx_0)\ =\ \int_{_{\NR^d}} 
\CP_n(t,\Nx+{\bf a};t_0,\Nx_0)\s\s da\ \equiv\ P_n(t,\Nx;t_0,\Nx_0)\s,
\label{redpr}
\qqq
with \s${\bf a}\equiv(a,\dots,a)\m$, \m gives the joint p.d.f. 
of the differences \s$x_{in}\s$ of the Lagrangian trajectories 
starting at points \s$\Nx_0\s$ (or, equivalently, the joint p.d.f.  
of the Lagrangian trajectories in the quasi-Lagrangian picture
\cite{BeLv}). It is translationally invariant separately
in \s$\Nx\s$ and \s$\Nx_0\m$. 
\vskip 0.3cm

In the limit \s$m\to0\s$ when 
\s$d^{\alpha\beta}(x)\s$ takes the scaling form (\ref{asy1})
but $\CD_0$ diverges, operator \s$M_n\m$, \m unlike \s$\CM_n\m$,
tends to the limit which for \s$\kappa=0\s$ coincides 
with the scaling operator 
\qq
M_n^{\rm sc}\ =\ \m\sum\limits_{1\leq i<j\leq n}\hs{-0.1cm}
d_{\rm sc}^{\alpha\beta}(x_{ij})\s\m\da_{x_i^\alpha}\da_{x_j^\beta}
\label{Mnsc}
\qqq
of scaling dimension \s$-\gamma\m$. \s$M^{\rm sc}_n\s$ is a 
positive singular elliptic differential operator of the 2$^{\rm\m nd}$ 
order in \s$L^2(\NR^{d_n})\m.$ \m By a simple self-consistent analysis 
one may convince oneself that, at least for \s$\gamma\s$
close to \s$2\m$, \s$\ee^{-t\m M_n}(\Nx,\Ny)\s$ and
\s$M_n^{\m-1}(\Nx,\Ny)\s$ converge pointwise when 
\s$m\to 0\s$ and \s$\kappa\to 0\s$ to the heat kernel 
\s$\ee^{-t\m M^{\rm sc}_n}(\Nx,\Ny)\s$ and the Green function 
\s$(M^{\rm sc}_n)^{-1}(\Nx,\Ny)\m$, \m respectively. The latter 
should satisfy bounds that may be inferred from a semi-classical 
analysis of the path integral expressions (\ref{expp1}) 
with \s$\CD^{\alpha\beta}\s$ replaced by \s$d_{\rm sc}^{\alpha\beta}\m$ 
and \s$\kappa\s$ set to zero\footnote{to our knowledge, such bounds 
have not been obtained in the mathematical literature and 
constitute an open mathematical problem}. In the limit \s$\gamma\to 
2\m$, \m \s$d_{\rm sc}^{\alpha\beta}(x)\s$ tends for non-zero 
\s$x\s$ to a constant times \s$\delta^{\alpha\beta}\s$ 
and \s$M_n^{\rm sc}\s$ becomes proportional 
to the \s$d_n$-dimensional Laplacian. When \s$\gamma\s$ is close
to \s$2\m$, the heat kernel 
\s$\ee^{-t\m M^{\rm sc}_n}(\Nx,\Ny)\s$
and the Green function \s$(M^{\rm sc}_n)^{-1}(\Nx,\Ny)\s$
differ little from the heat kernel and the Green function
of the Laplacian. In particular, \s$\ee^{-t\m M^{\rm sc}_n}(\Nx,\Ny)\s$
is finite everywhere and \s$(M^{\rm sc}_n)^{-1}(\Nx,\Ny)\s$
is infinite only when \s$\Nx=\Ny\m$. 
\m We expect this to hold for all \s$\gamma>0\m$.
\m When \s$2-\gamma\s$ is small, the behaviors 
of \s$(M^{\rm sc}_n)^{-1}(\Nx,\Ny)\s$
around \s$\Nx=\Ny\s$ and at infinity differ
from those of the Green function of the \s$d_n$-dimensional
Laplacian by \s$\CO({2-\gamma})\s$ modifications of the power laws. 
All that implies that the pointwise limits \s$m\to 0\s$ 
and \s$\kappa\to 0\s$ of the equal time correlators of \s$T\s$ given 
by Eqs.\s\s(\ref{2npf}) exist, at least for \s$\gamma\s$ close to 2,
and are given by the version of the same equations employing 
the scaling Green functions \s$(M^{\rm sc}_n)^{-1}(\Nx,\Ny)\s$
(with \s$d\Ny\s$ replaced by \s$d'\Ny\m$).
{}From now on, we shall deal only with these limits and
with the scaling operators \s$M^{\rm sc}_n\s$ and shall 
drop the superscript "sc".
\vskip 0.5cm

We are interested in the behavior of the equal time
correlators of \s$T\s$, especially in their scaling properties, 
in the situation when the source acts only on large distances,
i.e., in our Gaussian model, when the  spatial part
\s$\CC\s$ of the covariance of \s$f\s$ is almost constant. 
We may study this regime by replacing \s$\CC(x)\s$ by
\s$\CC_L(x)\equiv\CC(x/L)\s$ and by examining the large \s$L\s$
behavior of the equal time correlators \s$\CF_{2n}\equiv\CF_{2n,L}\m$. 
\m The following result, which may be referred to as
the {\bf zero mode dominance}, has been described in \cite{GK} and 
\cite{BGK}: 
\ at \s$\gamma\s$ sufficiently close to 2 and at \s$m,\kappa=0\m$,
\qq
\CF_{2n,L}(\Nx)\ =\ A_{\CC}\s L^{\rho_{2n}}\s\s \CF_{2n}^0(\Nx)
\ +\ \CO(L^{-2+\CO({2-\gamma})})\ +\ \s[\ .\ .\ .\ ]
\label{mres}
\qqq
for \s$n>1\s$. Above, \s$A_{\CC}\s$ is a non-universal 
amplitude (a constant
depending on the shape of the source covariance \s$\CC\m$) and
\m$\rho_{2n}\s=\s{2n(n-1)\over d+2}\m(2-\gamma)\m+\m\CO((2-\gamma)^2)\s$
is a universal (i.e.\s\s$\CC$-independent) anomalous exponent. 
\m$\CF_{2n}^0\s$ is the scaling (translationally 
invariant) zero mode of \s$M_{2n}\m$,
\qq
\CF_{2n}^0(\la\Nx)\s=\s\la^{\gamma\m n-\rho_{2n}}\s \CF_{2n}^0(\Nx)\s,
\quad\quad M_{2n}\s \CF_{2n}^0\s=\s 0\s.
\label{pres}
\qqq
$[\ .\ .\ .\ ]\s$ denotes terms which 
do not depend on at least one \s$x_i\s$ and as such do not
contribute to the correlation functions of scalar
differences \s$\langle\m\prod\limits_i(T(t,x_i)-T(t,y_i))\m\rangle\m$.
\qq
\CF_{2n}^0(\Nx)\ =\ \CS\ x_{12}^{\m2}x_{34}^{\m2}\s\cdots\s
x_{2n-1,\m2n}^{\s2}\ +\ \s\CO({2-\gamma})\ +\ [\ .\ .\ .\ ]
\qqq
where \s$\CS\s$ is the symmetrization operator. The 
contribution to \s$\CF_{2n}^0\s$ proportional to \s${2-\gamma}\s$
is also known up to \s$[\ .\ .\ .\ ]\s$ terms \cite{BGK}. 
A similar analysis was performed in \cite{Falk} and \cite{Falk2} 
for large space dimensions \s$d\m$.
\vskip 0.5cm

The main implication of the relation (\ref{mres}) is the
anomalous scaling of the \s$n>1\m$, \s$\gamma\s$ close 
to \s$2\s$ structure functions 
\s$S_{2n,L}(x)\equiv\langle\s(T(t,x)-T(t,0))^{2n}\m\rangle\m$. 
\m At \s$m,\kappa=0\s$ and for \s$\vert x\vert/L\ll 1\m$,
\qq
S_{2n,L}(x)\ \sim\ L^{\rho_{2n}}\s
\vert x\vert^{\gamma\m n-\rho_{2n}}\s.
\label{anos}
\qqq
The above behavior contradicts the simple dimensional prediction 
\s$S_{2n}(x)\s\sim\s\vert x\vert^{\gamma\m n}\s$ which holds only
for the 2-point function.
\vskip 0.5cm

Let us sketch the argument leading to the result (\ref{mres}),
based on applying the Mellin transform to select the dominant
contributions for large \s$L\m$. \m 
It will be convenient to work with a version of operators 
\s$M_n\s$ of scaling dimension zero  
\qq
N_n\ =\ R_n^{\gamma/2}M_n\m R_n^{\gamma/2}
\label{nn}
\qqq
where \s$R_n^2\equiv\sum_{_{i<j}}(x_i-x_j)^2\m$.
\s$N_n\s$ is also a positive (unbounded) 
operator\footnote{technically, \s$N_n\m$, \m as well as \s$M_n\m$,
may be defined as the Friedrichs extension of its restriction 
to smooth functions with compact support, vanishing around 
the diagonals \s$x_i=x_j\m$} in \s$L^2(\NR^{d_n})\m$.
\m Since it commutes with the self-adjoint
generator of dilations 
\qq
D_n\s=\s{_1\over^i}(\sum_ix_i^\alpha
\da_{x_i^\alpha}+{_{d_n}\over^2})\s,
\label{diin}
\qqq
it is partially diagonalized by the Mellin transform of 
the translationally invariant functions
\qq
f(\Nx)\ \ \ \rightarrow\ \ \ \hat 
f(\sigma,\hat{\Nx})\s=\s
\int\limits_0^\infty\la^{-\sigma-1}\s f(\la\hat{\Nx})\s\s d\la\s.
\label{Mell}
\qqq
The map (\ref{Mell}) is a unitary transformation, 
diagonalizing \s$D_n\m$, \m between $$L^2(\NR^{d_n})
\s\quad\ \ \ {\rm and}\ \ \ \quad\s L^2(\{{\rm Re}\s
\sigma=-{_{d_n}\over^2}\})\otimes
L^2(S^{d_n-1})\s,$$  
where \s$S^{d_n-1}\s$ is composed 
of points \s$\hat{\Nx}=\Nx/R_n\s$
in the space \s$\NR^{d_n}\s$ of difference variables. 
In the language of the Mellin transform, \s$N_n\s$ becomes
a family \s$\hat{N}_n(\sigma)\s$ of operators in \s$L^2(S^{d_n-1})\m$.
\m In particular,
\qq
(N_n^{\m-1}f)^{_{\widehat{{\ }}}}(\sigma,
\hat{\Nx})\ =\ \int\limits_{S^{d_n-1}}
\hat{N}_n^{\m-1}(\sigma;\m\hat{\Nx},\hat{\Ny})
\s\s\hat f(\sigma,\hat{\Ny})\s\s d\hat{\Ny}\s.
\qqq
where the Mellin-transformed Green function 
\s$\hat{N}_n^{\m-1}(\sigma;\m
\hat{\Nx},\hat{\Ny})\s$ satisfies the hermiticity relation
\qq
\overline{\hat{N}_n^{\m-1}(\sigma;\hat{\Ny},\hat{\Nx})}
\ =\ \hat{N}_n^{\m-1}(-d_n-\overline{\sigma};
\s\hat{\Nx},\hat{\Ny})\s.
\qqq
It is a meromorphic function of \s$\sigma\s$ with 
simple poles for generic \s$\gamma\m$. Around the poles
\qq
\hat{N}_n^{\m-1}(\sigma-{_\gamma\over^2};\s\hat{\Nx},\hat{\Ny})\ 
\ \cong\ \ {_1\over^{\sigma-\sigma_i}}\s\s\s f_i(\hat{\Nx})\m\ 
{\overline{g_i(\hat{\Ny})}}
\label{poles}
\qqq
where \s$f_i\s$ are the scaling zero modes of \s$M_n\s$
of scaling dimension \s$\sigma_i\s$ and \s$g_i\s$ are similar
modes with scaling dimensions \s$-d_n+\gamma-\overline{\sigma}_i\m$,
both in \s$L^2(S^{d_n-1})\m$. \m
Although operator \s$M_n\s$ has continuous spectrum
when considered as a positive operator in \s$L^2(\NR^{d_n})\m$,
\m it induces an operator \s$\hat{N}_n(\sigma-{_\gamma\over^2})\s$ 
in \s$L^2(S^{d_n-1})\s$ with a discrete spectrum 
when acting on scaling functions
with a scaling dimension \s$\sigma\m$. \m The scaling 
zero modes occur at discrete values \s$\sigma_i\s$ of \s$\sigma\s$
for which zero belongs to the spectrum.
\vskip 0.4cm

It is easy to see from the inductive equations (\ref{2npf}) that
\qq
\CF_{2n,L}(\Nx)\s=\s L^{n\m\gamma}\s\m\CF_{2n,\m1}(\Nx/L)
\qqq
and that, with the use of the Mellin transform, these equations
may be rewritten as
\qq
\CF_{2n,L}(\Nx)\ \ =\ \ L^{\gamma\m n}
\hs{-0.8cm}\int\limits_{{\rm Re}\s
\sigma=-{_{d_n}\over^2}+{_\gamma\over^2}}\hs{-0.8cm}{_{d\sigma}
\over^{2\pi i}}\ \s(R_n/L)^{\sigma}\s\ \hat{N}_n^{\m-1}
(\sigma-{_\gamma\over^2},\s\hat{\Nx},\hat{\Ny})\hs{1.6cm}\cr
\cdot\ (\CF_{2n-2,\m1}\otimes\CC)
^{_{\widehat{{\ }}}}(\sigma-\gamma,\s\hat{\Ny})\m\ d\hat{\Ny}\s.
\label{tocomp}
\qqq
Shifting the integration contour to \s${\rm Re}\s\sigma=\gamma\m n
+2-\CO({2-\gamma})\s$, we obtain
\qq
\CF_{2n,L}(\Nx)\ \ =\ \ -\sum\limits_{i}L^{\gamma\m n
-\sigma_i}\s\s 
R_n^{\sigma_i}\int{\rm Res}_{_{\sigma=\sigma_i}}\s\s
\hat{N}_n^{\m-1}(\sigma-{_\gamma\over^2};\s\hat{\Nx},\hat{\Ny})
\hs{1.6cm}\cr
\cdot\ (\CF_{2n-2,\m1}\otimes\CC)^{_{\widehat{{\ }}}}(\sigma-\gamma,
\s\hat{\Ny})\ d\hat{\Ny}\s\ +\ \ \CO(L^{-2+\CO({2-\gamma})})
\label{shc}
\qqq
where the sum runs over the poles \s$\sigma_i\s$ in the strip
\qq
-{_{d_n}\over^2}+{_\gamma\over^2}\s<\s{\rm Re}\s\sigma_i
\s<\s\gamma\m n+2-\CO({2-\gamma}) 
\label{strip}
\qqq
and the last term, suppressed 
for large \s$L\m$, \m comes from the shifted contour.
There are two types of poles: those coming from
\s$(\CF_{2n-2,\m1}\otimes\CC)^{_{\widehat{{\ }}}}\s$ and those
in the Green function \s$\hat{N}_n^{\m-1}\m$. \s The first 
ones contribute either to \s$[\ .\ .\ .\ ]\s$ or to
\s$\CO(L^{-2+\CO(2-\gamma)})\s$ in Eq.\s\s(\ref{mres}) and are not
interesting for us (at least for \s$\gamma\s$ close to 2). \m The 
second ones are related to the scaling zero modes 
of \s$M_n\m$, \m see Eq.\s\s(\ref{poles}). Only 
rotationally invariant (if \s$\CC\s$ 
has the same property) zero modes symmetric under
permutations of points and square-integrable 
on \s$S^{d_n-1}\s$ contribute to \s$\CF_{2n,L}\m$. \m 
Such zero modes may be studied for \s$\gamma\s$ close
to 2 by perturbative 
analysis of discrete-spectrum operators \s$M_n\s$ acting 
on scaling functions or, equivalently, of operators 
\s$\hat{N}_n(\sigma)\s$ acting in \s$L^2(S^{d_n-1})\s$.
(Recall that for \s$\gamma=2\m$, \s$M_n\s$ becomes
the \m$d_n$-dimensional Laplacian).
The result is that, for \s$\gamma\s$ close to 2, 
all but one zero modes 
in the strip (\ref{strip}) contribute \s$[\ .\ .\ .\ ]\s$
terms. \s$\CF_{2n}^0\s$ is the exception and it has scaling
dimension \s$\sigma_0=\gamma\m n-\rho_{2n}\m$. 
\m We expect essentially the same picture with the zero mode 
domination of correlation functions to persist for all
\s$\gamma>0\m$. \m One of the open problems is whether there 
are other \m non-$[\ .\ .\ .\ ]\s$  zero modes entering 
the strip (\ref{strip}) for smaller \s$\gamma\s$
and whether, if they cross, they may produce 
pairs of zero modes with complex scaling dimensions.
{}For \s$\gamma=0\s$ the singularities in the inverse
symbols of operators \s$M_n\s$ become strong enough to induce 
continuous spectrum of operators \s$\hat{N}_n(\sigma)\s$ 
and the picture of zero mode dominance has to be somewhat 
modified \cite{SS}\cite{BCKL}.
\vskip 0.5cm

One may also read the \s$\CO({2-\gamma})\s$ contribution
to the anomalous exponent \s$\rho_{2n}\s$ from 
the \s$\CO(({2-\gamma})\m\ln{L})\s$ term in the expansion 
of \s$\CF_{2n,L}\s$ into powers of \s${2-\gamma}\m$, 
\m similarly as in the \s$\epsilon$-expansion for critical 
phenomena one obtains anomalous exponents from 
logarithmic divergences. In the latter case, the renormalization
group which exponentiates the divergent logarithms
provides an explanation why it is reasonable to
extract information from badly divergent expansions.
In our argument, the Mellin transform analysis played 
a similar role exponentiating the logarithms of \s$L\s$.
One may show \cite{ca} that there is 
an (inverse) renormalization group picture of the advection 
problem hidden behind the above analogy. 
The renormalization group for the passive scalar
eliminates inductively the long-distance modes, unlike 
in critical phenomena where it is based on subsequent 
elimination of the short-distance degrees of freedom.
\vskip 0.8cm

\nsection{Conserved scaling structures}

In view of the domination of the 
equal time correlators of the scalar
by the scaling zero modes of operators
\s$M_n\m$, \m it is important to understand the physical
interpretation of such modes. It is, in fact, very simple:
\vskip 0.2cm

{\bf zero modes are scaling structures preserved in mean 
by the flow}.
\vskip 0.2cm

\noindent Indeed. Recall that \s$\ee^{-t\m M_n}
(\Nx,\Nx_0)\s=\s P_n(t,\Nx;0,\Nx_0)\m$ and it describes 
the probability that the differences of \s$n\s$ Lagrangian 
trajectories starting at time \s$0\s$ from points \s$\Nx_0\s$ 
are at time \s$t\s$ equal to \s$x_{in}\m$. The mean value 
of a translationally invariant function \s$f(\Nx)\s$
of positions of \s$n\s$ fluid particles at time \s$t\s$
is then equal to
\qq
\langle\s f\s\rangle_{_{t,\m{\bf x}_0}}\ \equiv\ 
\int f(\Nx)\s\s P_n(t,\Nx;0,\Nx_0)\s\s d'\Nx\ =\ 
\int f(\Nx)\s\s \ee^{-t\m M_n}(\Nx,\Nx_0)\s\s d'\Nx\s.
\label{mean}
\qqq
Differentiating the right hand side w.r.t. $t\m$,
\m we obtain
\qq
\int f(\Nx)\ M_n\s\ee^{-t\m M_n}(\Nx,\Nx_0)\s\s d'\Nx
\ =\ \int M_n f(\Nx)\ \ee^{-t\m M_n}(\Nx,\Nx_0)
\s\s d'\Nx
\label{inbp}
\qqq
where we have integrated by parts twice. If \s$f\s$
is a zero mode of \s$M_n\s$ then the right hand side vanishes
and, consequently, the mean (\ref{mean}) is constant and
\qq
\langle\s f\s\rangle_{_{t,\m{\bf x}_0}}\ =\ f(\Nx_0)\s.
\label{cons}
\qqq
\vskip 0.3cm

In fact, the story is a little bit more complicated. 
The zero modes with \s${\rm Re}\s\sigma_i>-d_n+\gamma\s$
are true zero modes. However the ones with the real part of their 
dimension \s$\leq-d_n+\gamma\s$ are not. For them, \s$M_n f\s$ 
is a contact term supported at the origin. Such contact terms may give 
non-zero contributions to the right hand side 
of Eq.\s\s(\ref{inbp}) or to the boundary terms
in the integration by parts, depending on the interpretation.
The zero modes with the scaling dimensions belonging to the strip 
(\ref{strip}) are true zero modes and hence they describe 
scaling structures of the flow conserved in mean. 
As was mentioned before, the translationally invariant zero modes
that are square-integrable on \s$S^{d_n-1}\s$ come in pairs 
\s$(f_i,g_i)\s$ corresponding to scaling dimensions \s$\sigma_i\s$ 
and \s$-d_n+\gamma-\overline{\sigma}_i\m$ (we may assume
that \s${\rm Re}\s\sigma_i\geq-{d_n\over2}+{\gamma\over 2}\m$).
\m For \s$\gamma\s$ close to \s$2\s$ there are no zero modes 
square integrable on \s$S^{d_n-1}\s$ with dimensions in the strip 
\s$-d_n+\gamma<{\rm Re}\s\sigma<0\m$. \m We expect this to hold
for any \s$\gamma>0\m$. \m In that situation \s$f_i\s$ are the true
zero modes and they have non-negative real parts of the scaling
dimension whereas \s$g_i\s$ are the false zero modes with real parts
of dimension \s$\leq -d_n+\gamma\s$ and with \s$M_ng_i\s$ being 
contact terms. Our claim about the absence of zero modes in the strip
\s$-d_n+\gamma<{\rm Re}\s\sigma<0\s$ may seem paradoxical
if we recall that the multi-body structure of operators \s$M_n\s$
assures that zero modes of \s$M_{n-p}\s$ are also annihilated by
\s$M_n\m$. Indeed\footnote{K.G. thanks E. Balkovsky, G. Falkovich 
and V. Lebedev for a discussion of this point}, the (false) zero modes 
of \s$M_{n-p}\s$ with \s${\rm Re}\s\sigma\leq-d_{n-p}+\gamma\s$ may 
lie in the strip \s$-{d_n\over^2}+{\gamma\over^2}<{\rm Re}\s\sigma<0\m$. 
\m However, the resulting zero modes of \s$M_n\s$ 
are not in \s$L^2(S^{d_n-1})\m$ and do not contribute 
to the poles of the Green function \s$\hat{N}^{-1}_n(\hat{\Nx},
\hat{\Ny})\s$ and hence to the right hand side of Eq.\s\s(\ref{shc}).
\vskip 0.5cm

It should be stressed that the behavior (\ref{cons}) is atypical.
{}For a general translationally invariant, scaling 
function with (say, positive) dimension \s$\sigma\s$
and for any time \s$\tau>0\m$,
\qq
\langle\s f\s\rangle_{_{t,\m{\bf x}_0}}\ 
=\ ({_t\over^\tau})^{^{\sigma\over\gamma}}
\int f(\Nx)\ \ee^{-\tau\m M_n}(\Nx,
\m({_\tau\over^t})^{^{1\over\gamma}}
\m\Nx_0)\s\s d'\Nx
\label{grwt}
\qqq
as it is easy to see with the use of the scaling property
\qq
\ee^{-\lambda^{^{\gamma}}\m t\m M_n}(\lambda\Nx,\lambda\Nx_0)
\ =\ \lambda^{-d_n}\s\s\ee^{-t\m M_n}(\Nx,\Nx_0)\s.
\label{prsc}
\qqq
It follows that, typically,
\qq
\langle\s f\s\rangle_{_{t,\m{\bf x}_0}}\ \ \sim\s\s\ \ t^{^{\sigma
\over\gamma}}\s.
\label{sdiff}
\qqq
The behavior (\ref{sdiff}) characterizes a {\bf super-diffusion}
where the square distances between points grow faster
than linearly in time. 
A slower behavior requires vanishing of
\s$\int f(\Nx)\s\s\ee^{-\tau\m M_n}(\Nx,0)\s\m d'\Nx\m$.
\m Note that the exponent of the growth
diverges when \s$\gamma\to 0\m$. 
\vskip 0.4cm

It is easy to understand the origin of the behavior 
(\ref{sdiff}). The stochastic process described by 
the probabilities \s$P_n(t,\Nx;0,\Nx_0)=\ee^{-t\m M_n}(\Nx,\Nx_0)\s$
may be viewed as a diffusion with the diffusion coefficient
proportional to the power $2-\gamma$ of the distance between 
the particles. When particles separate they diffuse
faster and faster which results in the super-diffusive behavior
with mean distance square growing proportionally 
to \s$t^{2/\gamma}$. \m On the other hand, on small distances
the diffusion is slow and particles which get close spend
relatively long time together. Since 
\s$\langle\s f\s\rangle_{_{0,\m{\bf x}_0}}=\s f(\Nx_0)\m$,
the time \s$t\s$ after which 
\s$\langle\s f\s\rangle_{_{t,\m{\bf x}_0}}\s$ reaches, say, 
twice its original value behaves like 
\s$\CO(f(\Nx_0)^{\gamma\over\sigma})\m$, i.e.\s\s it goes 
slower to zero with the diminishing separation between 
the initial points than for the standard diffusion at $\gamma=2$.
\vskip 0.45cm

We have seen that the scaling zero modes of \s$M_n\s$ correspond
to {\bf conserved collective modes} of the super-diffusion with 
the transition probabilities \s$P_n(t,\Nx;t_0,\Nx_0)\m$.
\m Existence of such conserved modes is nothing exceptional. 
They are present already in the standard diffusion. For example,
\qq
\int [\m x_{12}^{\s2}-x_{13}^{\s2}\m]\  
\ee^{\m t\m\Delta}(\Nx,\Nx_0)\s\s
d\Nx\ =\ x_{0,12}^{\s2}-x_{0,13}^{\s2}
\qqq
and is time independent, although \s$\int x_{ij}^{\s2}\s\m
\ee^{\m t\m\Delta}(\Nx,\Nx_0)\s\m d\Nx\s\m$
behaves like \s$\CO(t)\m$. \m Another example is
\qq
\int[\m x_{12}^{\s2}x_{34}^{\s2}
-\m{_d\over^{2(d+2)}}\s(x_{12}^{\s4}
+\m x_{34}^{\s4})]\s\s\ee^{\m t\m\Delta}(\Nx,\Nx_0)\s\s d\Nx
\s\m =\s\m x_{0,12}^{\s2}x_{0,34}^{\s2}-\m\s{_d\over^{2(d+2)}}\s
(x_{0,12}^{\s4}+\m x_{0,34}^{\s4})\s.\hs{0.4cm}
\qqq
Under symmetrization, the first conserved mode vanishes 
whereas the second one gives the zero mode 
of \s$\Delta\s$ whose \s$({2-\gamma})$-perturbation dominates
the 4-point function of the scalar for \s$\gamma\s$ close to 2.
\vskip 0.8cm

\nsection{Some physics: short-distance behavior 
of fluid particles}

The zero mode dominance of the structure functions
of the scalar is due to the appearance of such modes
in the asymptotics of the Green functions
\s$M_n^{-1}(\Nx,\Ny)\m$. \m Indeed, with the use of the
Mellin transform, one may write (in the reduced space):
\qq
M_n^{-1}(\Nx/L,\m\Ny)\ \s=\ \hs{-0.8cm}\int\limits_{{\rm Re}\s
\sigma=-{_{d_n}\over^2}+{_\gamma\over^2}}\hs{-0.8cm}{_{d\sigma}
\over^{2\pi i}}\ \s R_n(\Nx/L)^{\m\sigma}\s\ \hat{N}_n^{\m-1}
(\sigma-{_\gamma\over^2},\s\hat{\Nx},\hat{\Ny})\ 
R_n(\Ny)^{\m-{{d_n}}+\gamma-\sigma}\s,\hs{0.3cm}
\label{Melll}
\qqq
compare to Eq.\s\s(\ref{tocomp}). Pushing the integration
contour more and more to the right and using Eq.\s\s(\ref{poles}) 
to control the residues of the poles, we obtain for \s$\Ny\not=0\s$
the asymptotic large \m$L\m$ expansion:
\qq
M_n^{-1}(\Nx/L,\m\Ny)\ \s=\ \sum\limits_{i}\s
L^{-\sigma_i}\s\m f_i(\Nx)\s\s\overline{g_i(\Ny)}
\label{asgf}
\qqq
with \s${\rm Re}\s\sigma_i>-{{d_n}\over2}+{\gamma\over2}\m$ \m or,
as we expect, with \s${\rm Re}\s\sigma_i\geq 0\m$.
\m Although it has a similar form to the eigen-function expansion
of an operator with discrete spectrum, it has little to do
with the spectral decomposition of \s$M_n^{-1}$. \m Since \s$M_n\s$
is a positive operator in \s$L^2(\NR^{d_n})\s$ with continuous
spectrum coinciding with the positive real line, the spectral
decomposition of \s$M_n^{-1}\s$ is a continuous integral involving 
the generalized eigen-functions of \s$M_n\m$. \m  The scaling zero 
modes \s$f_i\s$ or \s$g_i\s$ of scaling dimensions
\s$\sigma_i\s$ and \s$-{{d_n}}+\gamma-\overline{\sigma}_i\m$,
\m respectively, are square-integrable on \s$S^{d_n-1}\s$
but are not generalized eigen-functions of \s$M_n\s$ (except for 
\s$\sigma_i=0\m$). They are rather analogous to resonances
in many-body problems with the plane of complex \s$\sigma\s$
replacing that of complex energies and \s$\sigma\s$ with 
real part equal to \s$-{d_n\over^2}\s$ corresponding to real 
energies\footnote{this analogy is somewhat loose, since
the poles in \s$\sigma\s$ live in the first sheet, probably
only on the real axis}. 
Note that due to the hermiticity and to the overall 
scaling of the Green function \s$M_n^{-1}(\Nx,\Ny)\m$,
\qq
M_n^{-1}(\lambda\Nx,\lambda\Ny)\ =\ \lambda^{\gamma-d_n}\m M_n^{-1}
(\Nx,\Ny)\s,
\label{scpp}
\qqq
the expansion (\ref{asgf}) may be also rewritten as 
\qq
M_n^{-1}(L\Nx,\Ny)\ =\ \sum\limits_{i}\s L^{-d_n+\gamma-
\overline{\sigma}_i}\ g_i(\Nx)\ \overline{f_i(\Ny)}
\label{oo}
\qqq
so that the zero modes \s$g_i\s$ with the scaling dimensions
\s$-d_n+\gamma-\overline{\sigma}_i\s$ of real part less than
\s$-{d_n\over 2}+{\gamma\over 2}\s$ (or even \m$\leq\m -d_n
+\gamma\m$) \m dominate the large
distance behavior of the Green function of \s$M_n\m$.
\m Expansion (\ref{oo}) may be also obtained directly from
Eq.\s\s(\ref{Melll}) by pushing the \s$\sigma$-integration 
contour to the left.
\vskip 0.4cm

It is not difficult to see directly that the functions
\s$f_i\s$ appearing in expansion (\ref{asgf}) have to describe
scaling structures conserved by the flow. Indeed,
\qq
\int M_n^{-1}(\Nx/L,\Ny)\s\s\ee^{-\tau\m M_n}(\Nx,\Nx_0)\s\s d'\Nx
\ =\ \sum\limits_{i}\s L^{-\sigma_i}\s\s\overline{g_i(\Ny)}
\int f_i(\Nx)\s\s\ee^{-\tau\m M_n}(\Nx,\Nx_0)\s\s d'\Nx
\label{as3}
\qqq
if we insert expansion (\ref{asgf}) into the left hand side. 
But on the other hand,
\qq
\da_\tau\int M_n^{-1}(\Nx/L,\Ny)\s\s\ee^{-\tau\m 
M_n}(\Nx,\Nx_0)\s\s d'\Nx
\ =\ \da_\tau\s\s L^{d_n-\gamma}\int M_n^{-1}(\Nx,L\Ny)\s\s
\ee^{-\tau\m M_n}(\Nx,\Nx_0)\s\s d'\Nx\cr
=\ -\m L^{d_n-\gamma}\s\s\ee^{-\tau\m M_n}(\Nx_0,\m L\Ny)
\hs{0.6cm}
\qqq
where we have used the scaling property (\ref{scpp}). The 
(reduced space) heat kernel on the right hand side
decays in \s$L\s$ faster 
than any power for \s$\Ny\not=0\m$. \m Comparing
the latter expression to
relation (\ref{as3}), we infer that \s$\da_\tau
\int f_i(\Nx)\s\s\ee^{-\tau\m M_n}(\Nx,\Nx_0)\s\s d'\Nx\s$
has to vanish and hence \s$f_i\m$, \m a function
with scaling dimension \s$\sigma_i\m$, \m is conserved in mean
by the Lagrangian flow. This gives another proof of the
statement (\ref{cons}).
\vskip 0.5cm

Since the Green function is given by the time integral of the
heat kernel,
\qq
M_n^{-1}(\Nx,\Ny)\ =\ \int\limits_0^\infty\ee^{- t\m M_n}
(\Nx,\Ny)\s\s dt\s,
\label{hcgf}
\qqq
one may also expect to see the zero modes in the asymptotic
behavior of the probabilities \s$P_n(t,\Nx;0,\Nx_0)
=\ee^{-t\m M_n}(\Nx,\Nx_0)\m$. \m Assume an asymptotic expansion
\qq
P_n(t,\Nx/L;\m 0,\Nx_0)\ =\ \sum\limits_{j}\s L^{-\rho_j}
\ \phi_j(\Nx)\ \m\overline{\psi_j(t,\Nx_0)}\s,
\label{aspt}
\qqq
with \s${\rm Re}\s\rho_j\geq0\m$, \m describing asymptotics 
of the probabilities that the Lagrangian
trajectories will come at time \s$t\s$ close together.
We could expect that functions \s$\phi_j\s$ are again zero modes 
of \s$M_n\m$. \m To verify whether this is the case, consider
the integral
\qq
\int P_n(t,\Nx/L;\m 0,\Ny)\s\s\ee^{-\tau\m M_n}(\Nx,\Nx_0)\s\s d'\Nx
\m=\m\sum\limits_{j}\m L^{-\rho_j}\s\m
\overline{\psi_j(t,\Ny)}\int\phi_j(\Nx)\s\m\ee^{-\tau\m M_n}
(\Nx,\Nx_0)\s\s d'\Nx\m.\hs{0.6cm}
\label{tc1}
\qqq
The left hand side may be rewritten as
\qq
L^{d_n}\int\ee^{-t\m M_n}(\Nx,\Ny)\ \ee^{-\tau\m M_n}(L\Nx,\Nx_0)
\s\s d'\Nx\ =\ \int\ee^{-t\m M_n}(\Nx,\Ny)\ \ee^{- L^{^{-\gamma}}\tau
\m M_n}(\Nx,\Nx_0/L)\s\s d'\Nx\cr 
=\ \ee^{-(t\m+\m L^{^{-\gamma}}\tau)\m M_n}(\Nx_0/L,\m\Ny)\ =\ 
\sum\limits_{j}\m L^{-\rho_j}\s\s
\phi_j(\Nx_0)\ \overline{\psi_j(t\m+\m L^{^{-\gamma}}\tau,\m\Ny)}
\hs{1cm}
\qqq
by changing variables \s$\Nx\mapsto L\Nx\s$ and using the scaling 
relations (\ref{prsc}), the composition law of heat kernels
and, finally, the expansion (\ref{aspt}).
But \s$\psi_j(t,\Ny)\s$ should be smooth in \s$t\s$ for
\s$t\not=0,\infty\m$. \s It then follows that
\qq
\int P_n(t,\Nx/L;\m 0,\Ny)\ \ee^{-\tau\m M_n}(\Nx,\Nx_0)\s\s d'\Nx
\ =\sum\limits_{j\atop p=0,1,\dots}\hs{-0.2cm}L^{-\rho_j-\gamma\m p}
\ {_{\tau^p}\over^{p!}}\ \da_t^{\m p}\m\overline{\psi_j(t,\Ny)}
\ \phi_j(\Nx_0)\s.\hs{0.7cm}
\label{tc2}
\qqq
The right hand side becomes independent of \s$\tau\s$ only 
approximately if \s$\tau L^{-\gamma}\m\ll\s 1\m$. 
Comparing the asymptotic expansions
(\ref{tc1}) and (\ref{tc2}), we infer that the
scaling functions \s$\phi_j\s$ (of scaling dimension \s$\rho_j\m$)
\m are not necessarily preserved in mean by the Lagrangian flow.
Instead, \s$\int\phi_j(\Nx)\s\m\ee^{-\tau\m M_n}(\Nx,\Nx_0)\s\m 
d'\Nx\s$ is a pure polynomial in \s$\tau\m$. \m Note
that the polynomial still grows
slower than the super-diffusive growth 
\s$\tau^{\m\rho_j/\gamma}\m$ since its order 
\s$p\s$ satisfies the relation 
\s$\rho_j=\rho_{j'}+\gamma\m p\geq\gamma\m p\s$
where \s$\rho_{j'}\s$ is a scaling dimension of
some other \s$\phi_{j'}\m$. \m Hence functions 
\s$\phi_j\s$ describe {\bf slow collective modes} 
of the super-diffusion. 
\vskip 0.3cm

It is not difficult to see how the slow modes are related 
to the zero modes of operators \s$M_n\m$. Differentiating
Eq.\s\s(\ref{aspt}) over \s$t\s$ we infer that
\qq
-\sum\limits_{j}\s L^{-\rho_j}\ \phi_j(\Nx)
\ \m\da_t\m\overline{\psi_j(t,\Nx_0)}&=& 
\sum\limits_{j}\s L^{-\rho_j}\ \phi_j(\Nx)
\ \m M_n\overline{\psi_j(t,\Nx_0)}\cr 
&=&\sum\limits_{j}\s L^{-\rho_j+\gamma}\ M_n\phi_j(\Nx)
\ \m\overline{\psi_j(t,\Nx_0)}\s.  
\label{diif}
\qqq
It follows that if the function \s$\phi_j\s$ appears 
in the expansion (\ref{aspt}) then also \s$M_n\phi_j\s$
does. Since the scaling dimension of \s$M_n\phi_j\s$ is 
\s$\rho_j-\gamma\s$ and only dimensions with real part positive
may appear, subsequent application of \s$M_n\s$ to a \s$\phi_j\s$
must produce a homogeneous zero mode of \s$M_n\s$ after
a finite number of steps. Hence functions \s$\phi_j\s$ must
be organized into towers of descendants \s$\phi_{i,p}\s$ 
based at zero modes \s$f_i\equiv\phi_{i,0}\s$ of \s$M_n\s$ 
and satisfying the chain of equations
\qq
M_n \m \phi_{i,p}\s=\s \phi_{i,p-1}\s, \quad \ \ p=1,\dots\s.
\label{chn}
\qqq
The scaling dimension of \s$\phi_{i,p}\s$ is $(\sigma_i+\gamma p)$.
Since \s$M_n^{p+1}\phi_{i,p}=0\m$, it follows that
the \s$(p+1)^{\rm\m th}\s$ time derivative of
\qq
\int\phi_{i,p}(\Nx)\ \ee^{-\m \tau M_n}(\Nx,\Nx_0)\s\s d'\Nx
\qqq
vanishes so that the above integral is a polynomial
in \s$\tau\s$ of degree \s$p\m$, in accordance with the
previous reasoning.
Note that Eq.\s\s(\ref{diif}) implies that
the functions \s$\psi_{i,p}\s$ corresponding to \s$\phi_{i,p}\s$
satisfy 
\qq
\psi_{i,p}\s=\s-\m\da_t\m\psi_{i,p-1}\s=\s M_n\m\psi_{i,p-1}\s.
\qqq
{\bf Summarizing}: \m the asymptotics of probabilities
of the Lagrangian trajectories to get close together is dominated
by the towers of slow collective modes of the super-diffusion:
\qq
P_n(t,\Nx/L;\m 0,\Nx_0)\ =\ \sum\limits_{i\atop p=0,1,\dots}\s 
L^{-\sigma_i-\gamma\m p}
\ \phi_{i,p}(\Nx)\ \m\overline{\psi_{i,p}(t,\Nx_0)}\s.
\label{aspt1}
\qqq
Since \s$P_n(t,\Nx;\m 0,\Nx_0)=P_n(t,\Nx_0;\m 0,\Nx)\m$,
expansion (\ref{aspt1}) may be also rewritten as
\qq
P_n(t,\Nx;\m 0,\Nx_0/L)\ =\ \sum\limits_{i\atop p=0,1,\dots}\s 
L^{-\sigma_i-\gamma\m p}
\ \m\overline{\psi_{i,p}(t,\Nx)}\ \phi_{i,p}(\Nx_0)\ \s.
\label{aspt2}
\qqq
giving the asymptotics of the probabilities of the Lagrangian 
trajectories starting very close. The leading term on
the right hand side is equal to \s$\overline{\psi_{0,0}(t,\Nx)}=
\ee^{-t\m M_n}(\Nx,0)\s$ and it corresponds to the constant
zero mode \s$\phi_{0,0}=f_0=1\m$.
\vskip 0.5cm

The above results have an important physical 
significance for the dynamics of the scalar.
Recall the expression (\ref{grwt}) for the time-dependence
of the mean value \s$\langle\s f\s\rangle_{_{t,\m{\bf x}_0}}\s$
of a function \s$f\s$ of scaling dimension \s$\sigma\m$.
Inserting the relation (\ref{aspt2}) to the
right hand side of Eq.\s\s(\ref{grwt}) we obtain
the asymptotic expansion of \s$\langle\s f\s\rangle_{_{t,
\m{\bf x}_0}}\s$ for large \s$t\s$:
\qq
\langle\s f\s\rangle_{_{t,\m{\bf x}_0}}\ 
=\ \sum\limits_{i\atop p=0,1,\dots}({_t\over^\tau})^{^{{\sigma
-\sigma_i\over\gamma}\m-p}}\ \phi_{i,p}(\Nx_0)\s\int f(\Nx)\s\s
{\overline{\psi_{i,p}(\tau,\Nx)}}\s\s d'\Nx\s.
\label{tosub}
\qqq
The leading term corresponds to the constant zero mode.
This term dominates for large \s$t\s$ if \s$\int f(\Nx)\s\m
\ee^{-\tau\m M_n}(\Nx,0)\s\m d'\Nx\not=0\m$. \m However 
if \s$f=\phi_{l,q}\s$ then, as we have seen 
above, \s$\langle\s \phi_{l,q}\s\rangle_{_{t,\m{\bf x}_0}}\s$
is a polynomial of order \s$q\s$ in \s$t\m$. Consequently,
\s$\int\phi_{l,q}(\Nx)\s\m
{\overline{\psi_{i,p}(\tau,\Nx)}}\s\m d'\Nx\s$ has to vanish 
unless \s${\sigma_l-\sigma_i\over\gamma}\m +q-p\s$ is an integer
between \s$0\s$ and \s$q\m$. \m Hence 
\s$\langle\s f\s\rangle_{_{t,\m{\bf x}_0}}\s$
for \s$f\s$ equal to a slow mode \s$\phi_{l,q}\s$ with a positive 
scaling dimension is dominated by
subleading terms on the right hand side of 
Eq.\s\s(\ref{tosub}). 
\vskip 0.3cm

To see the lower order terms\footnote{K.G. thanks M. Vergassola 
for the discussion of this point} for a generic scaling function
\s$f\s$ for which \s$\langle\s f\s\rangle_{_{t,\m{\bf x}_0}}
\s\sim\s t^{^{\sigma\over\gamma}}\m$, \m it is enough to compare 
the mean values \s$\langle\s f\s\rangle_{_{t,\m{\bf x}_0}}\s$ 
for two different \s$\Nx_0\m$. For example, subtracting 
\s$\langle\s f\s\rangle_{_{t,\m{\bf x}_0}}\s$
for two values of \s$x_{0,1}\s$ gets rid 
of the contribution of the constant mode. 
Denote by \s$\delta_{y_{0,m},y'_{0,m}}\s$ the operator
which performs the subtraction on functions
\s$h\s$ of \s$x_{0,m}\m$:
\qq
\delta_{y_{0,m},y'_{0,m}}\m h\s=\s h(y_{0,m})\m-\m h(y'_{0,m})\s.
\qqq
Subtracting subsequently at two different values 
of \s$x_{0,m}\s$ for \s$m=1,\dots,n-1\m$, \m
and setting \s$\prod\limits_{m=1}^{n-1}\delta_{y_{0,m},y'_{0,m}}
\equiv\delta_{\Ny,\Ny'}\m$, \m we obtain
\qq
\delta_{\Ny,\Ny'}\m
\langle\s f\s\rangle_{_{t,\s\cdot\s}}\ 
=\ \sum\limits_{i\atop p=0,1,\dots}{\hs{-0.25cm}'\hs{0.2cm}}
({_t\over^\tau})^{^{{\sigma-\sigma_i\over\gamma}\m-p}}\ 
(\delta_{\Ny,\Ny'}\m\phi_{i,p}\m)\s\m\int f(\Nx)\s\s
{\overline{\psi_{i,p}(\tau,\Nx)}}\s\s d'\Nx
\qqq
where the primed sum omits the contributions of the slow modes 
which do not depend on all variables. Hence the non-constant 
slow modes dominate the relative motion 
of groups of Lagrangian trajectories starting from different
initial configurations. In particular, the relative motion
is slower than the super-diffusive spread of the trajectories.
This supports the interpretation of the slow modes as 
{\bf resonance-type} objects in the motion of Lagrangian 
trajectories. The slow modes depending on less variables
correspond to resonances in fewer-particle channels
which drop out under the subtractions. 
\vskip 0.8cm

\nsection{Some mathematics: structure of the multi-body 
operators \s$M_n\m$}

To understand analytically the origin of the asymptotic expansion
(\ref{aspt1}), let us examine closer operators \s$M_n\m$.
We shall work in the reduced space \s$\CH\s\equiv\s 
L^2(\NR^{d_n})\m$. \s$M_n\s$ is a positive, unbounded, 
self-adjoint operator in \s$\CH\m$. \s Let 
\qq
(\CU_sf)(\Nx)\ =\ \ee^{\m s\m d_n/2}\s\s f(\ee^s\Nx)\s.
\qqq
Operators \s$\CU_s\s$ form a unitary version of the 1-parameter 
group of dilations in \s$\CH\s$ with the self-adjoint
operator \s$D_n\s$ of Eq.\s\s(\ref{diin}) as its generator:
\qq
\CU_s\ =\ \ee^{\m i\m s\m D_n}\s.
\qqq
\s$\CU_s\s$ preserve the domain of \s$M_n\s$ and
\qq
\CU_s\s M_n\s\m\CU_s^{-1}\ =\ \ee^{-\gamma\m s}\s\m M_n\s.
\label{crin}
\qqq
Denote by \s$X_n\s$ the natural logarithm of operator 
\s$M_n\s$: \ \s$X_n=\m\ln{M_n}\m$. \s$X_n\s$ is an unbounded 
self-adjoint operator on \s$\CH\s$ with the domain invariant
under \s$\CU_s\s$ and the whole real line as the spectrum. 
The relation (\ref{crin}) is equivalent to 
\qq
\CU_s\m X_n\m\CU_s^{-1}\ =\ X\s-\s\gamma\m s\s,
\qqq
i.e.\s\s to a strong form of the canonical commutation relation
\qq
[D_n,\m X_n]\ =\ i\s\gamma\s.
\qqq
Since under the Mellin transform (\ref{Mell}) \s$D_n\s$
becomes the multiplication operator by 
\s${1\over i}(\sigma+{d_n\over 2})\m$, 
\s$X_n\s$ must be unitarily equivalent to \s$\gamma\m\da_\sigma\s$ 
by virtue of the von Neumann Theorem on 
representations of the canonical commutation relations.
More exactly, there exists a one-parameter family 
\s$\hat U_n(\sigma)\m$,
\s${\rm Re}\s\sigma=-{{d_n}\over2}\m$, 
\m of unitary operators in \s$L^2(S^{d_n-1})\m$, 
\m unique modulo a right multiplication
by a \s$\sigma$-independent unitary operator, 
such that
\qq
(X_nf)^{_{\widehat{{\ }}}}(\sigma,\s\cdot\s\m)\ 
=\ \gamma\s\s\hat U_n(\sigma)\s\s\da_\sigma
\s\m\hat U_n^{-1}(\sigma)\s\m \hat{f}(\sigma,\s\cdot\s\m)
\qqq
for \s${\rm Re}\s\sigma=-{d_n\over 2}\m$. 
\m Equivalently, 
\qq
(M_nf)^{_{\widehat{{\ }}}}(\sigma,\s\cdot\s\m)\ 
=\ \hat U_n(\sigma)\s\s\ee^{\m\gamma\m\da_\sigma}
\s\s\hat U_n^{-1}(\sigma)\s\m \hat{f}(\sigma,\s\cdot\s\m)
\label{struc0}
\qqq
or, noting that the operator \s$\da_\sigma\s$ corresponds to 
the multiplication by \s$-\ln{R_n}\s$ in the language of the
original functions, 
\qq
M_n\ =\ U_n\m\s R_n^{-\gamma}\s\m 
U_n^{-1}
\label{struc}
\qqq
where 
\qq
(U_n f\m)^{_{\widehat{{\ }}}}
(\sigma,\s\cdot\s\m)\ =\ 
\hat U_n(\sigma)\m\hat f(\sigma,\s\cdot\s\m)\s.
\qqq
This is the promised structural result about operators \s$M_n\m$.
\m For the heat kernels, we obtain
\qq
\ee^{\m-t\m M_n}\ =\ U_n\s\s \ee^{\m -t\m R_n^{-\gamma}}\s\s 
U_n^{-1}\s.
\label{strhk}
\qqq
\vskip 0.4cm

What is the relation of the expression (\ref{struc})
to the representation 
\qq
M_n\ =\ R_n^{-\gamma/2}\s\s N_n\s\s R_n^{-\gamma/2}
\label{str2}
\qqq
used before, with \s$N_n\s$ an operator commuting with \s$D_n\m$?
The comparison of the two equations gives
\qq
N_n\ =\ R_n^{\m\gamma/2}\s\m U_n\m\s R_n^{-\gamma}\m\s U_n^{-1}
\m\s R_n^{\m\gamma/2}\ \ \quad{\rm or}\ \ \quad
N_n^{-1}\ =\ R_n^{-\gamma/2}\s\m U_n\m\s R_n^{\m\gamma}\m\s U_n^{-1}
\m\s R_n^{-\gamma/2}\s.
\label{tbrw}
\qqq
Let us suppose that the family of operators \s$\hat U_n(\sigma)\s$ 
has a meromorphic continuation to the complex plane of \s$\sigma\s$
with no poles in the strip 
\qq
-{_{d_n}\over^2}\leq{\rm Re}\s\sigma\leq
-{_{d_n}\over^2}+{_\gamma\over^2}
\qqq
(this will prove consistent with our zero mode analysis). 
Then \s$R_n^{-\gamma/2}\s\m U_n\m\s R_n^{\m\gamma/2}\s$
becomes under the Mellin transform the operator
$$\ee^{\m{_\gamma\over^2}\m\da_\sigma}\s\m\hat{U}_n(\sigma)
\s\m\ee^{-{_\gamma\over^2}\m\da_\sigma}\s=\s
\hat{U}_n(\sigma+{_\gamma\over^2})\s.$$
The unitarity of \s$\hat{U}_n\s$ for \s${\rm Re}\s\sigma=
-{d_n\over 2}\s$ implies that
\qq
\hat U_n(\sigma)\s \hat U_n(-d_n-\overline{\sigma})^*\s=\s
\hat U_n(-d_n-\overline{\sigma})^*\s\hat U_n(\sigma)\s=\s 1
\label{unit}
\qqq
and that \s$\hat U^{-1}(\sigma)\s$ also possesses a meromorphic 
continuation. Operator \s$R_n^{\m\gamma/2}\s\m U_n^{-1}
\m\s R_n^{-\gamma/2}\s$
becomes \s$\hat{U}_n^{-1}(\sigma-{_\gamma\over^2})\s$ 
under the Mellin transform and
Eq.\s\s(\ref{tbrw}) gives rise to the relation
\qq
\hat{N}_n^{-1}(\sigma-{_\gamma\over^2})\ =\ \hat U_n(\sigma)\  
\hat U_n^{-1}(\sigma-\gamma)\s.
\label{disre}
\qqq
\vskip 0.5cm

If \s$\hat{N}^{-1}_n(\sigma-{_\gamma\over^2})\s$ has a pole, 
see Eq.\s\s(\ref{poles}),
then the simplest possibility is 
that either \s$\hat U_n(\sigma_i-\gamma)\s$ 
is regular and then
\qq
\hat U_n(\sigma)\ =\ 
{_1\over^{\sigma-\sigma_i}}\ |f_i\rangle\langle g_i|\s 
\hat U_n(\sigma_i-\gamma)\ \ +\ \ \CO(1)
\label{1pos}
\qqq
when \s$\sigma\to\sigma_i\s$ or \s$ {\hat U}_n(\sigma_i)\s$
is regular and 
\qq
\hat U_n(\sigma_i)\ =\ 
{_1\over^{\sigma-\sigma_i}}\ |f_i\rangle\langle g_i|\ 
\hat U_n(\sigma-\gamma)\ \ +\ \ \CO(\sigma-\sigma_i)
\label{2pos}
\qqq
with \s$\langle g_i|\s\s\hat U_n(\sigma-\gamma)\s=
\s\CO(\sigma-\sigma_i)\m$. \s In the last case, multiplying
by \s$\hat U_n(-d_n-\overline{\sigma}_i)^*\s$
from the left and by \s$\hat U_n(-d_n-\overline{\sigma}+
\gamma)^*=\hat U^{-1}(\sigma-\gamma)\s$ 
from the right hand side and taking adjoints,
we obtain
\qq
\hat U_n(-d_n-\overline{\sigma}+\gamma)\ =\ 
{_1\over^{\overline{\sigma}-\overline{\sigma}_i}}\ |g_i\rangle
\langle f_i|\ \hat U_n(-d_n-\overline{\sigma}_i)\ \ 
+\ \ \CO(1)\s,
\qqq
i.e. relation (\ref{1pos}) for \s$\sigma_i\s$ replaced by
\s$-d_n+\gamma-\overline{\sigma}_i\m$ and corresponding to the
twin zero mode \s$g_i\s$ of \s$f_i\m$.
\m We expect the behavior (\ref{1pos}) if \s$f_i\s$ is
less singular at the origin than \s$g_i\s$ and the behavior (\ref{2pos})
in the opposite case (in Appendix A, this is established for 
\s$\hat U_2\m$). \m Rewrite Eq.\s\s(\ref{disre}) as
\qq
\hat U_n(\sigma+\gamma\m p)\ =\  
\hat{N}_n^{-1}(\sigma+\gamma(p-{_1\over^2}))\ 
\hat U_n(\sigma+\gamma(p-1))\s.
\label{mm}
\qqq
Assume that \s$\hat{N}_n^{-1}(\sigma_i+\gamma(p-{_1\over^2}))\s$
is regular for \s$p=1,2,\dots,\s$ (i.e. that there are no 
zero modes of \s$M_n\s$ (square-integrable on \s$S^{d_n-1}\m$)
with scaling dimensions differing by multiplicity of \s$\gamma\m$. 
\m This should hold for generic \s$\gamma\m$. \m From 
Eq.\s\s(\ref{mm}) for \s$p=0\s$ and from relation (\ref{1pos}) 
we infer that
\qq
\hat U_n(\sigma+\gamma)&=&
{_1\over^{\sigma-\sigma_i}}\ 
\hat{N}_n^{-1}(\sigma+{_\gamma\over^2})\ \vert f_i\rangle
\langle g_i\vert\ \hat U_n(\sigma_i-\gamma)\ \ +\ \ \CO(1)\cr\cr
&\equiv&{_1\over^{\sigma-\sigma_i}}\ 
\vert\phi_{i,1}\rangle \langle g_i\vert\ 
\hat U_n(\sigma_i-\gamma)\ \ +\ \ \CO(1)
\qqq
for \s$\sigma\to\sigma_i\m$. \m By induction on \s$p\m$, \m 
it follows then that
\qq
\hat U_n(\sigma+\gamma\m p))\ =\ {_1\over^{\sigma-\sigma_i}}
\ \vert\phi_{i,p}\rangle
\langle g_i\vert\ \hat U_n(\sigma_i-\gamma)\ \ +\ \ \CO(1)
\qqq
where
\qq
\phi_{i,p}\ =\ \hat{N}_n^{-1}(\sigma_i+\gamma(p-{_1\over^2}))\ 
\phi_{i,p-1}\s.
\label{chain}
\qqq
Note that Eqs.\s\s(\ref{chain}) may be rewritten
as the chain of relations (\ref{chn})
for the scaling functions \s$\phi_{i,p}(\Nx)\equiv R_n^{\sigma_i+
\gamma\m p}\s \phi_{i,p}(\hat{\Nx})\m$ where 
\s$\phi_{i,0}\equiv f_i\m$
is the zero mode of \s$M_n\m$ of scaling dimension
\s$\sigma_i\m$. \m By virtue of the assumption that 
there are no zero modes of \s$M_n\s$ of scaling dimension 
\s$\sigma_i+\gamma\m p\s$, the tower of descendants \s$\phi_{i,p}\s$ 
is uniquely determined\footnote{for non-generic \s$\gamma\s$ the
situation may be slightly more complicated with mixing of
different towers} for each zero mode \s$f_i\m$.
\vskip 0.5cm

%Note that functions \s$\phi_{i,p}\s$ satisfy the relation
%\qq
%\da_\tau^{\m p+1}\int \phi_{i,p}(\Nx)
%\ \ee^{-\tau\m M_n}(\Nx,\Nx_0)\ d'\Nx\ =\ 0
%\qqq
%and hence \s$\int
%\phi_{i,p}(\Nx)\s\s\ee^{-t\m M_n}(\Nx,\Nx_0)\s\s d'\Nx\s$
%is a polynomial in \s$\tau\s$ (of order \s$p\m$). \m Thus 
%functions \s$\phi_{i,p}\s$ have the same property as 
%functions \s$\phi_j\s$ appearing in the asymptotic expansion
%(\ref{aspt}). This, of course, is not a coincidence.
{}For \s$f\s$ a test function vanishing near the origin,
Eq.\s\s(\ref{strhk}) may be rewritten  as
%(for \s$f\s$ a test function vanishing near the origin) as
\qq
&&(\ee^{-t\m M_n} f)^{_{\widehat{{\ }}}}
(\sigma,\hat{\Nx})\cr
&&\hs{1cm}=\ 
\int d\hat{\Ny}\ \hat{U}_n(\sigma;\hat{\Nx},\hat{\Ny})
\int\limits_0^\infty d\lambda\s\s
\lambda^{-\sigma-1}\s\m\ee^{-t\m\lambda^{-\gamma}}\hs{-0.7cm}
\int\limits_{{\rm Re}\s\sigma'=-{d_n\over2}}\hs{-0.53cm}
{_{d\sigma'}\over^{2\pi i}}\s\s
\lambda^{\sigma'}\s\s(\m\hat{U}_n^{-1}\hat f\m)(\sigma',\hat{\Ny})
\s.\hs{0.3cm}
\qqq
After shifting the \s$\sigma'$-integration contour infinitesimaly 
to the left we may perform the \s$\lambda$-integral. 
The inverse Mellin transform of the resulting expression gives
\qq
&&(\ee^{-t\m M_n} f)(\hat{\Nx}/L)\cr \cr
&&\hs{0.9cm}=\ {_1\over^{\gamma}}\hs{-0.5cm}
\int\limits_{{\rm Re}\s\sigma=-{d_n\over2}}\hs{-0.53cm}
{_{d\sigma}\over^{2\pi i}}\ L^{-\sigma}
\int d\hat{\Ny}\ \hat{U}_n(\sigma;\hat{\Nx},\hat{\Ny})
\hs{-0.5cm}
\int\limits_{{\rm Re}\s\sigma'=-{d_n\over2}-0}\hs{-0.68cm}
{_{d\sigma'}\over^{2\pi i}}\ \m t^{^{\sigma'-\sigma\over\gamma}}\s\s
\Gamma({_{\sigma-\sigma'}\over^{\gamma}})
\ (\m\hat{U}_n^{-1}\hat f\m)(\sigma',\hat{\Ny})\s.\hs{1.2cm}
\label{topu}
\qqq
By moving the \s$\sigma$-integration contour further 
and further to the right\footnote{note that 
the poles of of the \m$\Gamma$-function do not contribute},
we obtain from Eq.\s\s(\ref{topu}) the asymptotic expansion
\qq
(\ee^{-t\m M_n} f)(\hat{\Nx}/L)\ =\ 
\sum\limits_{i\atop p=0,1,\dots}L^{-\sigma_i-\gamma\m p}
\s \phi_{i,p}(\hat{\Nx})\ \m{\overline{\psi_{i,p}}}
\label{asep}
\qqq
where the sum is over scaling dimensions
of zero modes of \s$M_n\s$ satisfying 
\s${\rm Re}\s\sigma_i>-{d_n\over2}\s$
and where
\qq
&&{\overline{\psi_{i,p}}}\ =\ -{_1\over^{\gamma}}
\int d\hat{\Ny}\ \overline{(\hat{U}_n(\sigma_i
-\gamma)^* g_i)(\hat{\Ny})}\cr
&&\hs{4.5cm}\cdot\hs{-0.5cm}\int\limits_{{\rm Re}
\s\sigma'=-{d_n\over2}}\hs{-0.53cm}
{_{d\sigma'}\over^{2\pi i}}\ 
\m t^{^{{\sigma'-\sigma_i\over\gamma}-p}}\s\s
\Gamma({_{\sigma_i-\sigma'}\over^{\gamma}}+p)
\ (\m\hat{U}_n^{-1}\hat f\m)(\sigma',\hat{\Ny})\s.\hs{0.5cm}
\qqq
Eq.\s\s(\ref{asep}) is an integrated 
version of expansion (\ref{aspt1})
(at least for \s$\gamma\s$ close to \s$2\m$, \m there are no 
scaling zero modes square-integrable on \s$S^{d_n-1}\s$ with
\s$-d_n<{\rm Re}\s\sigma_i<0\s$ and, as mentioned 
before, we expect this to hold for all positive \s$\gamma\m$).
\vskip 0.8cm

\nsection{Lagrangian flow}

There are some subtle points in the above discussion of the
motion of fluid particles. 
When stating that, for \s$m,\kappa=0\m$, \s$P_n(t,\Nx;0,\Nx_0)\s$ 
is the joint p.d.f.\s\s of the differences of the endpoints 
of \s$n\s$ Lagrangian trajectories, we have silently assumed 
that such trajectories, or at least their differences, make 
sense as random processes defining the Lagrangian flow 
on the probability space of \s$v\m$'s. \m A straightforward
consequence of such an assumption is the relation 
\qq
P_n(t,\Nx;0,\Nx_0)\bigg\vert_{_{x_{0,k}=x_{0,k+1}=\dots=x_{0,n}}}=
\quad P_{k}(t,\Nx';0,\Nx'_0)\ \prod\limits_{i=k}^n\delta(x_{in})
\label{tobe}
\qqq
where \s$\Nx'=(x_1,\dots,x_k)\s$ and similarly
for \s$\Nx'_0\m$. \m Eq.\s\s(\ref{tobe}) expresses the
elementary property that the joint p.d.f.\s\s of coinciding 
random variables is concentrated on the diagonal. 
In particular, \s$P_n(t,\Nx;0,0)\s$ should be proportional 
to the \s$d_n$-dimensional delta-function.
But the heat kernels \s$\ee^{-t\m M_n}(\Nx,\Nx_0)\s$ 
{\bf do not have} this property at least for \s$\gamma\s$
close to \s$2\s$ and, expectedly, for all \s$\gamma>0\m$.  
\m Instead they are regular when \s$\Nx_0\to0\m$. 
How exactly \s$P_n(t,\Nx;0,\Nx_0/L)\s$ fails to
become the delta-function when \s$L\s$ goes to infinity
is described by the asymptotics (\ref{aspt2}) dominated
by the slow collective modes of the stochastic evolution of
the Lagrangian trajectories. Hence, even if all 
joint p.d.f.'s \s$P_n\s$ of the differences 
of Lagrangian trajectories make sense as given by the 
\s$m,\kappa=0\s$ heat kernels \s$\ee^{-t\m M_n}
(\Nx,\Nx_0)\m$, \m the differences of Lagrangian trajectories 
do not exist as random processes for \s$\gamma>0\m$. 
Note that for \s$\kappa\s$ positive we should not expect
the behavior (\ref{tobe}) since the Brownian motions
starting from \s$x_{0,i}\s$ are different for 
different \s$i$'s even if they wiggle around the same 
Lagrangian trajectory. The system behaves as if the wigglings 
were present even for \s$\kappa=0\s$ (see more 
on that below).
\vskip 0.5cm

The \s$\gamma=0\m$ and \s$m,\kappa=0\m$ case will be analyzed in
Sects.\s\s8 and 9 and was previously considered in
\cite{SS2} to \cite{GKo}, see also \cite{SS1}. 
The p.d.f. $P_2(t,x;0,x_{0})=\ee^{-t\m M_2}(x,x_0)\s$ 
of the difference \s$x_{12}(t)\equiv x(t)\s$ of two 
Lagrangian trajectories may be easily computed in this case 
and the result is the log-normal distribution
\cite{CFKL}\cite{CGK}
\qq
P_2(t,x;0,x_0)
\ =\ {_{r^{-d}}\over^{\sqrt{4\pi D\m t}}}\ 
\ee^{\m-{1\over
4 D\m t}\s(\m\ln{r\over^{r_0}}\s-\s t\m Dd\m)^2}\ 
k_{D{_{d+1}\over^{d-1}}\m t}(\hat x,\hat x_0)
\label{tker}
\qqq
where \s$r=\vert x\vert\m$, \s$\hat x=x/r\s$
and similarly for \s$r_0\m,\ \hat x_0\m$. \m
\s$k_t(\hat x,\hat x_0)\s$ denotes the heat kernel
on the unit sphere in \s$d\s$ dimensions and it drops 
out in the rotationally invariant sector. The most
important consequence of Eq.\s\s(\ref{tker}) is 
that \s${1\over t}\m\ln{r\over r_0}\s$ is a Gaussian variable
with covariance \s$2D/t\s$ tending to zero at large times and 
with mean \s$Dd\m$. \m The mean gives the {\bf Lyapunov exponent}
i.e.\s\s the rate of exponential growth in time of the distance 
\s$r\s$ between the Lagrangian trajectories. Note that
\qq
\int r^{\sigma}\s\s P_2(t,x;0,x_0)\s\s
dx\ =\ r_0^{\sigma}\s\m\ee^{\m D\sigma(d+\sigma)
\s t}
\qqq
which should be contrasted with the super-diffusive 
behavior for \s$\gamma>0\m$ described by Eq.\s\s(\ref{sdiff}). 
\m More generally,
\qq
\int\hs{-0.08cm} f(x)\s\s P_2(t,x;0,x_0)
\s\s dx\s=\s\int\hs{-0.08cm} 
f(\ee^u\m r_0\m\hat x)\s\s {_1\over^{\sqrt{4\pi 
D\m t}}}\s\s\ee^{\m-{1\over4D\m t}\s(\m u\s-\s t\m Dd\m)^2}\s\s 
k_{D{_{d+1}\over^{d-1}}\m t}(\hat x,\hat x_0)\s\s du\s\s 
d\hat x\s.\hs{0.4cm}
\qqq
{}For any test function \s$f\s$ and for fixed \s$t\m$, \m the right 
hand side tends to \s$f(0)\s$ when \s$r_0\to0\m$,
\m in accordance with the relation (\ref{tobe}) and unlike 
for \s$\gamma>0\m$. As it is easy to see from the above integral 
(or from Eq.\s\s(\ref{tker})\m), 
\m the concentration of the p.d.f. $P_2(t,x;0,x_0)\s$ 
within \s$r\s{\mathop{<}\limits_{^\sim}}\s\eta\s$
is visible if \s$\eta\gg\ee^{tDd}\m r_0\m$. \m For the later use,
note that for a small but non-zero \s$r_0\s$ and for a rotationally 
invariant test function \s$f\m$,
\qq
\int f(x)\s\s P_2(t,x;0,x_0)
\s\s dx\ =\ \int f(\ee^{tu}\m r_0)\s\s {_{\sqrt{t}}
\over^{\sqrt{4\pi D}}}
\s\s\ee^{\m-{t\over4D}\s(\m u\s-\s Dd\m)^2}\s\s du
\quad\mathop{\rightarrow}\limits_{t\to\infty}\quad 0\s.
\label{insta}
\qqq
\vskip 0.4cm   
   
The approach of \cite{CFKL}\cite{CGK} was based on the 
observation that at \s$\gamma=0\s$ the 2-point function of the velocity 
differences becomes
\qq
\langle\s(v^\alpha(t_1,x_1)-v^\alpha(t_1,x_2))\s\m
(v^\beta(t_2,x'_1)-v^\beta(t_2,x'_2))\s\rangle\ =\ 
2\m {_D\over^{d-1}}\s\delta(t_{12})\hs{1.3cm}\cr\cr 
\cdot\ [\m(d+1)\s\delta^{\alpha\beta}\s x_{12}\cdot\m x'_{12}
\s-\s x^\alpha_{12}{x'}^\beta_{12}\s-\s{x'}^\alpha_{12}\m
x_{12}^\beta\m]\s.
\label{diffc}
\qqq
In particular, the first derivatives 
of \s$v\s$ have space-independent correlations. In other words, 
we may set 
\qq
v(t,x_1)-v(t,x_2)\s=\s X(t)\s\s x_{12}\quad{\rm or}\quad\da_\gamma 
v^\alpha(t,x)\s=\s X^{\alpha\gamma}(t)
\label{eqpro}
\qqq
where \s$X^{\alpha\beta}(t)\s$ is a Gaussian process
with values in traceless matrices with mean zero and 
the 2-point function
\qq
\langle\s X^{\gamma\alpha}(t)\s X^{\delta\beta}(s)\m\rangle
\ =\ 2\m {_D\over^{d-1}}
\s\delta(t-s)\s\s[\s(d+1)\s\delta^{\alpha\beta}
\m\delta^{\gamma\delta}\s-\s\delta^{\alpha\gamma}
\m\delta^{\beta\delta}\s-\s\delta^{\alpha\delta}
\m\delta^{\beta\gamma}\s]
\label{tracl}
\qqq
obtained by differentiating twice the right hand side of 
(\ref{diffc}). It is easy to check directly that the above 
covariance is positive and that it is invariant under the adjoint
action of \s$O(d)\m$, \m i.e. that \s$X\s$ and \s$kXk^{-1}\s$ 
have the same covariance for \s$k\in O(d)\m$. \m
Eqs.\s\s(\ref{eqpro}) are equalities between
Gaussian processes. Physically, they mean that for \s$m=0\s$
and \s$\gamma=0\s$ the velocity flow acts as a uniform, volume 
preserving strain and rotation, as far as the relative
motions of fluid particles are concerned. 
The difference of two Lagrangian trajectories 
\s$x_{12}(t)\equiv x(t)\s$ should satisfy
the linear (stochastic) ODE
\qq
dx\s=\s X(t)\s x\s dt\s,\quad\quad x(0)=x_0
\qqq
with a solution given formally by
\qq
x(t)\s=\s g_{t,t_0}\s\m x_0
\label{sso}
\qqq
where \s$g_{t,t_0}\s$ is the time-ordered exponential of
an integral of independent matrices,
\qq
g_{t,t_0}\s=\s \CT\s\s\ee^{\m\int_{t_0}^tX(s)\s ds}\s,
\label{tord}
\qqq
of the type similar to the ones that appears
in the theorems on products of independent equally
distributed matrices \cite{Furst} or in the 
one-dimensional Anderson localization \cite{Past}.
The point is that \s$g_{t,t_0}\s$ may be defined
as a random Markov process (a diffusion) 
with values in \s$SL(d)\m$. \m It has 
three basic properties:
\qq
&&\hs{-1.6cm}1.\quad g_{t_2,t_1}\s{\rm\ and\ \s}g_{t_2+\tau,t_1+\tau}\ \ 
{\rm have\ the\ same\ distribution},\cr
&&\hs{-1.6cm}2.\quad g_{t_2,t_1}\s g_{t_1,t_0}\s=\s g_{t_2,t_0}\ \ 
{\rm a.\m e.},\cr
&&\hs{-1.6cm}3.\quad g_{t,t_0}\ \ \ {\rm is\ independent\ of}\ \ \ 
g_{t',t'_0}\ \ \ {\rm if}\ \ \ (t_0,\m t)\cap(t'_0,\m t')\s
=\s\emptyset\s.
\nonumber
\qqq
To define such a process, it is enough to give the
(transition) probability distributions 
\s$p_{t-t_0}(g)\s dg\s$ of $g_{t,t_0}\s$
(\m$dg\s$ denotes the Haar measure on \s$SL(d)\m$)
satisfying the composition law:
\qq
\int p_{t}(g)\s\s p_{s}(g^{-1}h)\s\s dg\ =\ p_{t+s}(h)\s.
\label{sgl}
\qqq
The $SO(d)$-invariance of the Lie-algebra-valued process \s$X\s$ 
imposes also the relation 
\qq
p_t(kgk^{-1})\s=\s p_t(g)\s.
\label{oinv}
\qqq
In Sect.\s\s8 we identify \s$p_t\s$ with the heat kernel
of a certain operator on \s$SL(d)\m$.
\vskip 0.5cm

The net outcome of that analysis is 
that for \s$\gamma=0\m$, \m unlike for 
\s$\gamma>0\m$, \m the differences of Lagrangian trajectories 
\s$x_{ij}(t)=g_{t,t_0}\m x_{0,ij}\s$ are well defined random 
variables. \m In particular, the knowledge of \s$p_t\s$
is all what is needed to compute the joint p.d.f.'s 
of \s$x_{ij}(t)\s$:
\qq
\int P_n(t,\Nx;0,\Nx_0)\s\s f(\Nx)\s\s d'\Nx\ =\ 
\int\limits_{SL(d)}p_{t}(g)\s\s f(g\Nx_0)\s\s dg
\label{ksi2}
\qqq
for translationally invariant \s$f\m$. \m In fact,
the above integrals uniquely determine \s$p_t\s$.
\vskip 0.4cm

One of the consequences of the relation (\ref{ksi2}), 
closely related to the property
(\ref{tobe}), is that, for \s$\gamma=0\m$, the stochastic evolution
of the scalar \s$T\s$ defined by the \s$m,\kappa=0$ flow preserves 
the Gibbs measure formally given as
\s${\ee^{-\beta\int T^2}\s\m DT\over{\rm normalization}}\m$.
\m Indeed, the \s$2n$-point function of the scalar in this measure is 
\qq
\CF_{n}^{^{\rm Gibbs}}(\Nx)\ =\ {_{(2n)!}\over^{2^{2n}n!\m\beta^n}}
\s\s\CS\s\s\delta(x_{12})\m\delta(x_{34})\s\cdots\s\delta(x_{2n-1,2n})
\qqq
(the odd functions vanish). But Eq.\s\s(\ref{ksi2}) implies
the relation
\qq
\int P_{2n}(t,\Nx;0,\Nx_0)\s\s 
\CF_{n}^{^{\rm Gibbs}}(\Nx_0)\m\s\s d'\Nx_0
\ =\ \CF_{n}^{^{\rm Gibbs}}(\Nx)
\label{topl}
\qqq
i.e. the time invariance of the Gibbs measure correlations for
\s$\gamma=0\m$. \m This should be contrasted with the behavior
for the \s$\gamma>0\s$ case where the
flux of the scalar energy towards 
high wavenumbers destroys the invariance
of the Gibbs measure, see \cite{Schlad}. 
{}For \s$\gamma=0\m$, the invariant Gibbs measure 
is nevertheless unstable under perturbations, 
as follows from relation (\ref{insta}). It has also
little to do with the \s$\kappa\to0\s$ limit
of the stationary state of the scalar obtained in the presence
of large scale forcing. The latter 
will be constructed in Sect.\s\s9.
\vskip 0.5cm

The mathematics of the difference between 
the \s$\gamma=0\s$ and \s$\gamma>0\s$
cases is simple. Eq.\s\s(\ref{lagr}) requires that \s$v(t,x)\s$ be 
Lipschitz in \s$x\s$ for the uniqueness of solutions\footnote{recall 
the existence of two solutions with vanishing initial condition: 
\s$x=({_{\gamma}\over^2}\m t)^{^{2\over\gamma}}\m$ and \s$x=0\m$, 
\m for the equation \s$^{\cdot}\hs{-0.16cm}x=x^{{2-\gamma}\over2}\m$}. 
\m But the Gaussian \s$v$-measure with 2-point function 
(\ref{cv2}) lives on \s$v\m$'s which 
are H\"{older} in \s$x\s$ with exponent 
\s$({2-\gamma})/2\s$ (modulo logarithmic corrections) but not
Lipschitz, except for \s$\gamma=0\s$ where the velocity differences
become smooth, as we have seen above. Hence, one should
not expect uniqueness of Lagrangian trajectories even
if the probabilistic description of them may be maintained
but with violation of the property (\ref{tobe}). 
Physically\footnote{K.G. thanks G. Falkovich for
a discussion of this point}, the velocity covariance should be 
smoothed on the dissipative scale \s$\eta\s$
due to viscous effects so that it behaves 
as \s$\sim\m D\m\eta^{-\gamma}\s r^2\s$ for
\s$r\ll\eta\m$, \m i.e.\s\s like the \s$\gamma=0\s$ covariance
with \s$D\s$ increased to \s$D\m\eta^{-\gamma}\m$. \m 
The Lagrangian trajectories diverge 
now exponentially in time as long as their distance is \s$\ll\eta\m$. 
\m Note, however, that for arbitrary small but fixed 
\s$r_0=\vert x_{0}\vert\s$ one never sees concentration 
of the p.d.f. \s$P_2(t,x;0,x_0)\s$ on scales smaller 
than \s$\eta\s$ if \s$\eta\s{\mathop{<}\limits_{^\sim}}\s
\ee^{\m\CO(t\m\eta^{-\gamma})}\m r_0\m$, 
\m i.e. for \s$\eta\s$ sufficiently small.
This explains in more physical terms why relation (\ref{tobe}) 
fails when \s$\eta\to0$ for \s$\gamma>0\m$. \m The exponential 
divergence of trajectories closer than \s$\eta\s$ makes it 
impossible to maintain the concept of (differences of) 
individual trajectories in the inviscid limit \s$\eta\to 0\m$. 
\m Instead, we should talk about the \s$v$-dependent p.d.f. 
\s$P_n(t,\Nx;0,\Nx_0\m|\m v)\s$
whose average over the velocity ensemble reproduces 
\s$P_n(t,\Nx;0,\Nx_0)\m$. \m It is worth noting
that for positive diffusivity \s$\kappa\m$,
when the deterministic equation (\ref{lagr}) should be replaced
by the stochastic ODE (\ref{stoch}), although the problems with
the non-uniqueness of the solutions persist, there
exists a rigorous probabilistic treatment\footnote{we
thank G. Eyink for pointing this out to us} allowing to
define uniquely the transition probabilities 
\s$P_n(t,\Nx;0,\Nx_0\m|\m v)\s$ for H\"{o}lder
continuous velocities \cite{SV}. Our analysis calls for an
extension of such a treatment to the \s$\kappa=0\s$ case.
\vskip 0.8cm

\nsection{Advection by smooth velocities and harmonic\break analysis}

When \s$\gamma=0\m$ and \s$m,\kappa=0\m$, \m the Kraichnan model
becomes exactly solvable as we will show now. That simplifications
occur in this case has been noted before, see 
\cite{SS} and \cite{SS1} to \cite{FKLM}. Our analysis 
is based on some observations by Shraiman and Siggia \cite{SS}\cite{SS2}.
As was noted in \cite{SS} and in \cite{SS2}, for $\gamma=0$ the model
has extra symmetries. The operators \s$M_n\s$ can be expressed
in terms of the quadratic Casimir operators corresponding to
an action of the groups \s$SL(d)\s$ and \s$SO(d)\s$ on the
correlation functions. Let us explain what this means.
\vskip 0.5cm

The group \s$SL(d)\s$ of real matrices of determinant 1 acts on
functions \s$f\s$ on \s${\bf R}^d\s$ on the left by 
\s$(L_gf)(x)=f(g^{-1}x)\m$. \m
The infinitesimal form of this action is 
given by \s${d\over dt}|_{_{t=0}}
L_{e^{tA}}f=A^{\beta\alpha}H_{\alpha\beta}f\s$ where 
\s$A\s$ is a traceless matrix (i.e. in the Lie algebra of \s$SL(d)\m$)
\m and the generators \s$H_{\alpha\beta}\s$ are
\qq
H_{\alpha\beta}\s=\s- x^\alpha\da_{x^\beta}\m+\m{_1\over^d}\s
\delta_{\alpha\beta}\s x^\gamma\da_{x^\gamma}\s. 
\qqq
Similarily, on functions of \s$n\s$ \s${\bf R}^d\s$ variables 
\s$(x_1,\dots,x_n)=\bf x\m$, \m we have the (diagonal) action 
\qq
(L_gf)({\bf x})=f(g^{-1}{\bf x})\label{action}
\qqq
with generators 
$H_{\alpha\beta}\s=\s \sum_i (-\m x_i^\alpha\da_{x_i^\beta}
\m+\m{_1\over^d}\s
\delta_{\alpha\beta}\s x_i^\gamma\da_{x_i^\gamma})$\s. 
The quadratic Casimir of $SL(d)$ is in terms of these generators
\qq
H^2\ =\ \sum\limits_{\alpha,\beta}
H_{\alpha\beta}H_{\beta\alpha}\ .
\qqq
The generators of the action of the \s$SO(d)\s$ subgroup 
are \s$J_{\alpha\beta}=H_{\alpha\beta}-H_{\beta\alpha}\s$ 
and the corresponding
quadratic Casimir is
\qq
J^2\ =\ -\m{_1\over^2}\sum\limits_{\alpha,\beta}
J_{\alpha\beta}^{\s2}\s.
\qqq
The observation of Shraiman and Siggia was that when $\gamma,m=0$
in the velocity covariance $d^{\alpha\beta}$ 
(\ref{asy1}), the operator ${M}_n
= \sum_{i<j}d^{\alpha\beta}(x_i-x_j)\da_{x^\alpha_i}\da_{x^\beta_j}$
becomes
\qq
M_n\m=\m{_D\over^{d-1}}\m[\m(d+1)\m J^2\s-\s d\m H^2]\s.
\label{tbc0}
\qqq
By definition of the action (\ref{action}), the same
formula holds also when we express ${M}_n$ in terms 
of the $n-1$ difference variables. In particular
for \s$M_2\m$, \m the Casimirs \s$H^2\s$ and \s$J^2\s$ correspond
to the action of \s$SL(d)\s$ and \s$SO(d)\s$ on functions 
\s$f(x)\s$ of the difference variable \s$x\equiv x_{12}\m$.
\m In this case, we may diagonalize \s$H^2\s$ 
by the Mellin transform
\s$f(x)\s\rightarrow\s\hat f(\sigma,\hat{x})\m=
\m\int\limits_0^\infty r^{-\sigma-1}\s 
f(r\m\hat{x})\s\s dr\s\m$:
\qq
(H^2f)^{_{\widehat{{\ }}}}(\sigma,\hat{x})
\s=\s{_{d-1}\over^d}\m\sigma\m(\sigma+d)
\s\s\s \hat f(\sigma,\hat x)\s.
\label{tbcl}
\qqq
with \s${\rm Re}\s\sigma=-{d\over 2}\m$. \m It follows, 
in particular, that the spectrum \s$H^2\s$ acting in \s$L^2(\NR^d)\s$
is \s$]-\infty,\m-{d(d-1)\over^4}]\s$ and that of \s$M_2\s$
is \s$[{D\m d^2\over 4},\infty[\m$. \m Denoting 
\s$\ee^{-t\m J^2}\equiv k_t\m$, we obtain
\qq
&&\hs{-1cm}\int f(x)\s\s\ee^{-t\m M_2}(x,x_0)\s\s dx\cr\cr
&&=\ \int\limits_{{\rm Re}\s\sigma=-{_d\over^2}}\hs{-0.3cm}
{_{d\sigma}\over^{2\pi i}}\int\limits_0^\infty
dr\int d\hat{x}\ r^{-\sigma-1}\s\s f(r\hat{x})\ r_0^{\sigma}\s\s
\ee^{\m tD\m\sigma(\sigma+d)}\s\s 
k_{D{_{d+1}\over^{d-1}}\m t}(\hat{x},
\hat{x}_0)\cr
&&=\ \int\limits_0^\infty{_{dr}\over^{r}}\int d\hat{x}
\ f(r\hat{x})
\ {_1\over^{\sqrt{4\pi D\m t}}}\s\s\ee^{\m-{_1\over^{4D\m t}}
\s(\m \ln{{_r\over^{r_0}}}\s-\s t\m Dd\m)^2}\s\s
k_{D{_{d+1}\over^{d-1}}\m t}(\hat{x},\hat{x}_0)
\label{aa}
\qqq
where we have performed the Gaussian integral over
\s$\sigma\m$. The result (\ref{tker}) readily follows.
Taking \s$f(x)=\CC_L(r)=\CC(r/L)\s$ with \s$\CC\s$ 
the rotationally
invariant forcing covariance, we obtain by integrating 
over \s$t\s$ the expression for the 2-point function of 
\s$T\s$ at \s$\gamma=0\s$:
\qq
\CF_{2,L}(x)\s=\s(M_2^{-1}\CC_L)(x)\s=\s{_1\over^{Dd}}
\left(\int_{_r}^{^\infty}\CC(\rho/L)\s {_{d\rho}\over^\rho}
\s+\s r^{-d}\int_{_0}^{^r}\CC(\rho/L)\s \rho^{d-1}\m d\rho
\right).
\label{ff2p}
\qqq
Clearly \s$\CF_{2,L}(x)\s$ is smooth for \s$x\not= 0\s$ and
\qq
&&\CF_{2,L}(x)\s\cong\hbox to 3.2cm{$\s-\m{{\CC(0)}
\over{Dd}}\m\ln{(r/L)}$\hfill}
{\rm for\ small\ \s}r\s,\label{za1}\\
&&\CF_{2,L}(x)\s\sim\hbox to 3.2cm{$\s (r/L)^{-d}$\hfill}
{\rm for\ large\ \s}r\s.
\label{za2}
\qqq
\vskip 0.5cm

In order to solve Eqs.\s\s(\ref{2npf}) for the higher-point functions
of \s$T\s$ we need a representation
for the Green function \s${M}_n^{-1}\m$.
\m This is obtained by relating \s$H^2\s$ and \s$J^2\s$ to the Casimirs 
\s$\CH^2\s$ and \s$\CJ^2\s$ of the
left action of \s$SL(d)\s$ and of \s$SO(d)\s$ 
on functions \s$F\s$ on \s$SL(d)\m$, 
\m given by \s$(\CL_gF)(h)=F(g^{-1}h)\m$, \m or in the infinitesimal
form by \s${d\over dt}|_{_{t=0}}\CL_{e^{tA}}F
= A^{\beta\alpha}\CH_{\alpha\beta}F\m$. \m Note that
\s$\CH_{\alpha\beta}\s$ are skew-adjoint in the regular 
representation and that 
\qq
(d+1)\m\CJ^2\s-\s d\m\CH^2\ =\ -\m{_{d+2}\over^4}
\sum\limits_{\alpha,\beta}(\CH_{\alpha\beta}-\CH_{\beta\alpha})^2
\s-\s{_d\over^4}\sum\limits_{\alpha,\beta}
(\CH_{\alpha\beta}+\CH_{\beta\alpha})^2
\qqq
is a positive elliptic operator in \s$L^2(dg)\m$.
\m In particular, it has the heat kernel
\qq
\CK_t(g,h)\ \equiv\ \ee^{-t{D\over d-1}[(d+1)\CJ^2
-d\CH^2]}(g,h)
\label{heker}
\qqq
satisfying \s$\int\CK_t(g,h)\s dh=1\s$ and
\s$\CK_t(g,h)=\CK_t(kg,\m kh)=
\CK_t(gg',\m hg')\s$ for \s$k\in SO(d)\s$ and \s$g'\in SL(d)\m$.
\m Assign to a translationally invariant function 
\s$f(\Nx)\s$ and to \s$\Nx\s$ 
a function \s$F_\Nx(g)=f(g\Nx)\m$ on \s$SL(d)\m$.
\m The linear map \s$f\mapsto F\s$ intertwines 
the two actions of \s$SL(d)\m$:
\qq
(L_{g}f)(h\Nx)\s=\s(\CL_{g}F_\Nx)(h)\s.
\qqq
It follows that 
\qq
\int\ee^{-t\m M_n}(g\Nx,\Ny)\s\m f(\Ny)\s\s d'\Ny\ 
=\ \int\CK_t(g,h)\s\m f(h\Nx)\s\s dh\s.
\label{heat}
\qqq
Comparing the above relation to Eq.\s\s(\ref{ksi2})
we conclude that
\qq
p_t(hg^{-1})\ =\ \CK_t(g,h)\s.
\qqq
Clearly the basic properties (\ref{sgl}) and (\ref{oinv})
of the p.d.f. \s$p_t\s$ follow. In other words, we may identify
\s$g_{t.t_0}\s$ as the diffusion process on  
group \s$SL(d)\s$ with the generator equal to 
\s${D\over d-1}[(d+1)\CJ^2-d\CH^2]\m$.
\vskip 0.5cm 

Integrating the relation (\ref{heat}) over \s$t\s$ we infer that
\qq
\int M_n^{-1}(g\Nx,\Ny)\s f(\Ny)\s\s d'\Ny\ =\ \int\CG(g,h)\s\m 
f(h\Nx)\s\s dh
\label{green}
\qqq
where \s$\CG\s$ is the integral kernel of
\s$({_D\over^{d-1}}\m[\m(d+1)\m\CJ^2\s-\s d\m \CH^2])^{-1}\m$.
\m Applying iteratively identity (\ref{green}) to
Eq.\s\s(\ref{2npf}) we end up with the expression
\qq
\CF_{2n}(\Nx)=\sum_p F_{2n}(\Nu_p)
\label{IDN}
\qqq
where the sum runs through all ordered pairings 
\m$p=(\{i_1,j_1\},\dots,\{i_n,j_n\})\m$ of 
\m$\{1\dots 2n\}\m$, \m$\Nu_p=(x_{i_1j_1},\dots,x_{i_nj_n})\m$ and
\qq 
{}F_{2n}(\Nu)
\s=\s\int
\prod_{i=1}^n\CG(g_{i-1},g_i)\s\m\CC(g_iu_i)\s\m dg_i
\s=\s\int
\prod_{i=1}^n\tilde\CG(g_{i-1},g_i)\s\m\CC(g_iu_i)\s\m dg_i
\label{F2n}
\qqq
where \s$g_0=e\m$ and 
\s$\tilde{\CG}(g,h)=\int_{_{SO(d)}}\CG(g,kh)\s dk\m$.
The last equality follows by substituting
\s$g_1=k_1g'_1,\s g_2=k_1k_2g'_2,\s g_n=k_1\cdots k_ng'_n\m$,
\m and using 
\s$\CG(kg_1,kg_2)=\CG(g_1,g_2)\s$ and \s$\CC(kx)=\CC(x)\m$.
\vskip 0.5cm

The final reduction consists of identifying \s$\tilde{\CG}\s$
with the Green function of the Laplace-Beltrami operator \s$\Delta\s$ 
on the homogeneous space \s$H_d\equiv SL(d)/SO(d)\m$. 
\m By definition, \s$\Delta\s$ coincides with the Casimir 
\s$\hf\CH^2\s$ if we view functions on \s$H_d\s$ as functions 
on \s$SL(d)\s$ right-invariant under the action of \s$SO(d)\m$. 
\m Assign to a function \s$f\s$
on \s$H_d\s$ the function \s$g\mapsto\tilde f(g)=f(g^{-1})\m$.
\m Clearly \s$\tilde f(kg)\s=\s \tilde f(g)\m$ and
\s$(\CL_gf)(h^{-1})\s=\s\tilde f(hg)\s\equiv\s(\CR_g\tilde f)(h)\m$, 
\m i.e.\s\s\m the map \s$f\mapsto\tilde f\s$ intertwines the
action of \s$SL(d)\s$ on the functions on \s$H_d\s$ with
the right regular action of \s$SL(d)\m$. \m Since the quadratic 
Casimirs of \s$SL(d)\s$ in the left-regular 
and in the right-regular representations coincide and
\s$\CJ^2\s$ vanishes in the action on \s$\tilde f\m$, 
\m we infer that 
\qq
-2d(\Delta f)(g^{-1})\s=\s
-d(\CH^2 \tilde{f})(g)\s=\s([(d+1)\CJ^2-d\CH^2]\tilde f)(g)
\qqq
and that
\qq
\int G(g^{-1},\m h)\s\m f(h)\s\s dh\ =\ \int\CG(g,h)\s\m\tilde f(h)
\s\s dh\ =\ \int\tilde\CG(g,h)\s\m\tilde f(h)\s\s dh
\qqq
where the function \s$G(g,h)\s$ on \s$H_d\times H_d\s$ represents 
the kernel of the operator \s$(-D'\Delta)^{-1}\s$ where
\s$D'\equiv{2Dd\over d-1}\m$. 
\m Thus \s$\tilde{\CG}(g,h)=G(g^{-1},\m h^{-1})\s$ and 
Eq.\s\s(\ref{F2n}) becomes
\qq 
{}F_{2n}(\Nu)
\s=\s\int
\prod_{i=1}^n G(g_{i-1},g_i)\s\m\CC(g_i^{-1}u_i)\s\m dg_i
\label{F2n1}
\qqq
\vskip 0.4cm

Every matrix $g\in SL(d)$ can be uniquely represented
as a product (the so called Iwasawa decomposition)
\s$g=nak\s$ where \s$k\in SO(d)\m$, \s$n\s$ is upper triangular with
\s$1\s$ on the diagonal and \s$a\s$ is diagonal with positive
entries. Thus one may parametrize the cosets \s$gSO(d)\s$
by \s$na\m$. \m For \s$d=2\s$ we may write 
\s$a= {\rm diag}(y^{1\over 2},y^{-{1\over 2}})\m$, \s$y>0\m$, 
\s$n=(\matrix{_1&_x\cr^0&^1})\m$, 
\s$x\in \NR\m$. \m The Haar measure \s$dg\s$ becomes
\s$dg=y^{-2}\m dx\m dy\s dk\m$.
The homogeneous space \s$H_2\s$ may be identified with
the upper half-plane \s$H=\{z=x+iy\in \NC\s |\s y>0\}\m$. 
\m The action of \s$SL(2)\s$ 
on \s$H\s$ is given by the M\"obius transformations 
\s$(\matrix{_a&_b\cr ^c&^d})z
={az+b\over cz+d}\m$. \m Since \s$ki=i\m$, the identification
maps the coset \s$gSO(2)\s$ to \s$gi=nai\m$. \m We shall denote
\s$na\equiv g(z)=
(\matrix{_{y^{{1\over 2}}}&_{y^{-{1\over 2}}x}
\cr ^0&^{y^{-{1\over 2}}}})\m$. \m The \s$SL(2)$-invariant
measure on \s$H\s$ is \s$d\nu(z)=y^{-2}dx\m dy\s$ and 
the Laplace-Beltrami operator becomes
\qq
\Delta=y^2(\p_y^2+\p_x^2)\s.
\qqq
The Green function \s$G\s$ is given by the explicit expression:
\qq
G(z,z')={_1\over^{16D\pi}}\s\ln{(x-x')^2+(y+y')^2
\over (x-x')^2+(y-y')^2} \label{lap}.
\qqq
Eq.\s\s(\ref{F2n1}) may now be rewritten as
\qq 
{}F_{2n}(\Nu)
\s=\s\int
\prod_{i=1}^n G(z_{i-1},z_i)\s\m\CC(g(z_i)^{-1}u_i)\s\m d\nu(z_i)
\label{F2n2}
\qqq
with \s$z_0=i\m$.
In Appendix B we study the integrals (\ref{F2n2}) in more detail.
In particular we show that the leading singularities at
coinciding points of the correlation functions of \s$T\s$ 
are given by a Gaussian expression, a sum of
products of 2-point functions, confirming the analysis 
of \cite{CFKL}\cite{BCKL}.
\vskip 0.3cm

For the dimension \s$d>2\s$ 
one can proceed analogously. In the Iwasawa
decomposition we parametrize \s$n\s$ by the off-diagonal entries, 
\s$x_\alpha\m$, \s$\alpha=1,\dots, {d^2-d\over 2}\m$, \m and write 
\s$a=\ee^{\phi}\m$, \s$\phi={\rm diag}(\phi_1,\dots,\phi_d)\s$
with \s$\sum_i\phi_i=0\m$. \m The
Haar measure becomes in these variables
\qq
dg=\ee^{\m\sum_{i<j}
(\phi_j-\phi_i)}\prod_{i=1}^{d-1}d\phi_i\prod_\alpha 
dx_\alpha\s dk\s.
\qqq
\s$G\s$ is (proportional to) the Green function of the 
Laplace-Beltrami operator \s$\Delta\s$ on
\s$SL(d)/SO(d)\m$. \m Explicitly, 
for \s$d=3\s$ write \s$\phi=\hf\m\alpha\s{\rm
diag}(1,-1,0)+\m{_1\over^6}\m\beta\s{\rm
diag}(1,1,-2)\m$. \m Then
\qq
\Delta=\ee^{2\alpha}\p_{x_1}^2+\ee^{\alpha+\beta}\p_{x_2}^2+
\ee^{\beta-\alpha}(\p_{x_3}+x_1\p_{x_2})^2+\p_\alpha^2+3\p_\beta^2 .
\label{lb}
\qqq
and \s$dg=\ee^{-\alpha-\beta}d\alpha\s 
d\beta\s dx_1\s dx_2\s dx_3\s dk\m$.
\m There does not seem to exist 
a very explicit expression for \s$G\s$
in \s$d>2\m$. \m However, the singular 
behavior of \s$\CF_{2n}\s$ can
be extracted again, see Appendix B.
\vskip 0.5cm

Let us end this section by deriving the formula for 
the Lyapunov exponents of the Lagrangian trajectories, 
previously found in \cite{GKo} by path-integral techniques.
In \cite{SS} it was observed that \s$M_n\s$ may be also 
expressed using the quadratic Casimir of the action of
\s$SL(n-1)\s$ with the generators 
\qq
G_{ij}\s=\s -\m x^\alpha_{in}\m\da_{x_i^\alpha}
\s+\s{_1\over^{n-1}}\delta_{ij}\m\s x^\alpha_{kn}
\m\da_{x_k^\alpha}
\qqq
for  $1\leq i,j\leq n-1\m$. This action corresponds
to the natural action of \s$SL(n-1)\s$ on the
\s$i$-index of \s$x_{in}\equiv x_i-x_n\m$. \m Denoting by 
\s$G^2\s$ the quadratic Casimir \s$\sum_{_{i,j}}\hs{-0.1cm}
G_{ij}G_{ji}\s$
and by \s$\Lambda\s$ the generator of dilations \s$x_i^\alpha
\m\da_{x_i^\alpha}\m$, \m one obtains \cite{SS}
\qq
M_n\ =\ {_D\over^{d-1}}\s\m[\s
(d+1)\m J^2\s-d\m G^2\s-
\s{_{d-n+1}\over^{n-1}}\s
\Lambda\m(\Lambda+d_n\m)\s]\s.
\label{SS0}
\qqq
Let \s$\rho$ denote the volume spanned by vectors \s$x_{in}\m$, 
$i=1,\dots,n-1\m$, \m describing the time \s$t$ differences 
of the Lagrangian trajectories starting at time zero
from points \s$\Nx_0\s\m$:
\qq
\rho\s=\s\sqrt{{\det}_{_{i,j}}{(x_{in}\cdot\s x_{jn}\m)}}\s.
\qqq
We would like to find the p.d.f. of \s$\rho\m$. \m Note that
for a function \s$f(\rho)\m$,
\qq
M_n f(\rho)=-\m{_{(d-n+1)\m D}\over^{(d-1)(n-1)}}\s
\Lambda\m(\Lambda+d_n)\s\m f(\rho)=
-\m{_{(n-1)\m(d-n+1)\m D}\over^{d-1}}
\s \rho\m\da_\rho(\rho\m\da_\rho+d)\s\m f(\rho)\m.\hs{0.9cm} 
\qqq
This follows from Eq.\s\s(\ref{SS0}) 
since \s$\rho\s$ is \s$SL(n-1)$- 
and \s$SO(d)$-invariant. Hence \s$M_n\s$ preserves  
the space of functions \s$f(\rho)\m$. \m Also
\qq
\int\vert f(\rho)\vert^2\s\s\prod\limits_i dx_{in}
\ =\ {\rm const}.\s\int_{_0}^{^\infty}
\vert f(\rho)\vert^2\s \rho^{d-1} d\rho
\qqq
where \s${\rm const}.=\int\delta(\rho-1)\s\prod_{_i}dx_{in}\m$.
\m Hence \s$M_n\s$ in the action on \s$f(\rho)\s$ is diagonalized
by the Mellin transform
\ $f(\rho)\s\rightarrow\s\hat f(\sigma)\m=
\m\int\limits_0^\infty \rho^{-\sigma-1}\s 
f(\rho)\s\s d\rho\ $ (unitary for 
\s${\rm Re}\s\sigma=-{d\over 2}\m$)\m:
\qq
(M_n f)^{_{\widehat{{\ }}}}(\sigma)\ =\ 
-\m{_{(n-1)\m(d-n+1)\m D}\over^{d-1}}\s\s\sigma(\sigma+d)\s\s
\hat f(\sigma)\s\equiv\s-\m D(n)\s
\sigma(\sigma+d)\s\s\hat f(\sigma)\s.
\qqq
As in Eq.\s\s(\ref{aa}), we obtain
\qq
\int f(\rho)\s\s P_n(t,\Nx;0,\Nx_0)\s\s d'\Nx\ =\ 
\int\limits_0^\infty{_{d\rho}\over^\rho}\s\s f(\rho)\ {_1
\over^{\sqrt{4\pi D(n)\m t}}}\ \ee^{\m -\m{_1\over^{4D(n)\m t}}
\s(\ln{{_\rho\over^{\rho_0}}}\m-\m t\m D(n) d)^2}\s.
\label{gpdf}
\qqq
Hence \s${1\over t}\m\ln{\rho}\s$ is a Gaussian variable with
covariance \s$2D(n)/t\s$ tending to zero at large times
and with mean \s$D(n)d\s$ which, by definition, is the sum of the 
\s$(n-1)\s$ largest Lyapunov exponents describing the effective 
separation of \s$n\s$ Lagrangian trajectories. We infer that the 
\s$n^{\rm \m th}\s$ Lyapunov exponent is
\qq
\lambda_n\ =\ (D(n+1)-D(n))\m d\ =\ \hf\m (d-2n+1)\m D'
\qqq
with \s$d\s$ exponents equally spaced and symmetric with
respect to the origin, confirming the result of \cite{GKo}.
\vskip 0.8cm

\nsection{Quadrature of the \m$\gamma=0\m$ case}

Let us explicitly construct the stationary state of 
the passive scalar advected by smooth Gaussian velocity 
with 2-point function (\ref{diffc}).
Relations (\ref{IDN}) and (\ref{F2n1}) allow 
to write a compact expression
for the generating function of the $\gamma=0$
theory:
\qq
\Phi(\chi)\ \equiv\ 
\langle\s\ee^{\m i\int \chi(x)\m T(x)\m dx}\s
\rangle\s=\s\sum_{n=0}^{\infty}
{_{(-1)^n}\over^{(2n)!}}
\sum_p \int F_{2n}(\Nu_p)\prod_i \chi(x_i)\s\m
dx_i\s.
\label{quad0}
\qqq
Noting that all pairings give the same contribution to the
\s$x_i\s$ integral 
and that there are \s${(2n)!\over 2^n}\s$
of them we get
\qq
\Phi(\chi)
\s=\s\sum_{n=0}^{\infty}
(-1)^n\int\prod_{i=1}^nG(g_{i-1},g_i)\s V_\chi(g_i)\s
\m dg_i
\label{fex}
\qqq
where
\qq
V_\chi(g)=\hf\int\CC(g^{-1}(x-y))\s\s\chi(x)\s\chi(y)\s\s dx\s dy
\label{Vchi}
\qqq
is a non-negative function on \s$H_d\s$ bounded by
\s$V_\chi(e)\m=\m\hf\s \CC(0)\s(\int\chi)^2\m$. 
\m Eq.\s\s(\ref{fex}) may be rewritten in the operator language 
as (\m$D'\equiv{2Dd\over d-1}\m$)
\qq
\Phi(\chi)\s=\s\sum_{n=0}^{\infty}
(-1)^n\int[\m({_1\over^{\m-D'\Delta\m}}\m V_\chi)^n1\m](e)\s.
\qqq
The sum on the right hand side involves the Neuman series 
for the operator
\s$(-D'\Delta\m+\m V_\chi)^{-1}\m$, \m i.e.\m\s\s for
the Laplacian on \s$H_d\s$ perturbed by a potential.
Resumming the series we obtain
\qq
\Phi(\chi)
\ =\ 1\s-\s[\m(-D'\Delta\m+\m V_\chi)^{-1}V_\chi\m](e)
\label{quad}
\qqq
which is an explicit expression for the characteristic functional
of the stationary state of the \s$\gamma=0\s$ Kraichnan model.
\vskip 0.5cm

Let us see that the right hand side of Eq.\s\s(\ref{quad}) makes sense. 
Using the Feynman-Kac formula expressing the perturbed heat kernel 
as an expectation \s$E_g(\s\cdot\s)\s$ 
with respect to the Brownian motion on \s$H_d\s$ with transition
amplitudes \s$\ee^{-D'\Delta}(g,h)\m$, \m starting at time 
zero at \s$g\s$:
\qq
\ee^{-t\m(-D'\Delta\m+\m V_\chi)}(g,h)\ =\ E_g\left(\ee^{-
\int_0^t V_\chi(h(s))\m\s ds}\s\s\delta_h(h(t))\right),
\label{Brbr}
\qqq
we infer the bounds
\qq
&&0\s\leq\s\ee^{-t\m(-D'\Delta\m+\m V_\chi)}(g,h)\s\leq\s 
\ee^{\m t\m D'\Delta}(g,h)\s,\label{hhker}\\
&&0\s\leq\s(-D'\Delta\m+\m V_\chi)^{-1}(g,h)\s\leq\s 
G(g,h)\s.
\label{hgree}
\qqq
Since 
\qq
[\m(-D'\Delta\m+\m V_\chi)^{-1}V_\chi\m](e)\s=\s
\int(-D'\Delta\m+\m V_\chi)^{-1}(e,h)\s\m V_\chi(h)
\s\s dh\s,
\qqq
it follows that the latter integral is bounded 
by the smeared 2-point function
\qq
\int G(e,h)\s\m V_\chi(h)\s\s dh\s=\s \hf\int\CF_2(x_{12})
\m\s\chi(x_1)\s\chi(x_2)\s\m dx_1\s dx_2\s
\qqq
which is finite for test functions \s$\chi\s$ e.g.\s\s from
the Schwartz space \s$\CS(\NR^d)\m$, see Eq.\s\s(\ref{ff2p}).
\vskip 0.5cm

\s$\Phi\s$ defines a continuous positive-definite
functional on \s$\CS(\NR^d)\m$. \m The continuity 
of \s$\Phi(\chi)\s$ w.r.t. $\chi\in
\CS(\NR^d)\s$ is easy: it follows by the Dominated Convergence
Theorem from the Feynman-Kac representation of the perturbed 
Green function:
\qq
(-D'\Delta\m+\m V_\chi)^{-1}(g,h)\ =\ \int_{_0}^{^\infty}\hs{-0.2cm}dt
\s\s\s
E_g\left(\ee^{-
\int_0^{t}V_\chi(h(s))\s\m ds}\s\s\delta_h(h(t))\right).
\qqq
The positive definiteness:
\qq
\sum\limits_{r,s}\lambda_r\m\overline{\lambda_s}\s\s
\Phi(\chi_r-\chi_s)\ \geq\ 0\s,
\qqq
is a little bit more complicated. Let us sketch its proof.
Define first the positive definite characteristic functional 
\qq
\Phi_t(\chi)\ =\ \langle\s\ee^{\m i\int T(t,x)\s\chi(x)\s dx}
\s\rangle                            
\label{phit}
\qqq
of the time \s$t\s$ (quasi-Lagrangian) state of the scalar 
where \s$T(t,x)=\int_{0}^{t} f(s,\m g_{t,s}^{\s-1}x)\s ds\s$
is a functional of the forcing \s$f\s$ and of \s$g_{t,s}\m$. 
\m The above expression for \s$T(t,x)\s$ is obtained 
for the initial condition vanishing 
at \s$t_0=0\s$ in Eq.\s\s(\ref{solu}). \m The expectation 
in (\ref{phit}) is w.r.t.\s\s\m the Gaussian measure of the forcing 
and w.r.t.\s\s\m the measure of the diffusion process \s$g_{t,s}\m$. 
\m It is easy to see that \s$\int T(t,x)\s\chi(x)\s dx\s$
is square-integrable with respect to these measures. Performing 
the integration with respect to \s$f\m$, \m we obtain
\qq 
\Phi_t(\chi)\ =\ \langle\s\ee^{\m-\int_0^tV_\chi(g_{t,s})\m\s ds} 
\s\rangle\s.
\label{nnm}
\qqq
The remaining expectation over \s$g_{t,s}\s$ is easy to calculate
by expanding the exponential (the resulting series of expectations 
converges absolutely for finite \s$t\m$). \m The result is
\qq
\Phi_t(\chi)\ =\ 
1\s-\s\int_{_0}^{^t}
[\m\ee^{-s\m(-D'\Delta\m+\m V_\chi)}\m
V_\chi\m](e)\s\s ds\s.
\label{ane}
\qqq
Using the bound (\ref{hhker}), it is easy to see that
\s$\Phi_t(\chi)\s$ converge to \s$\Phi(\chi)\s$ when \s$t\to\infty\m$.
\m Hence the positive definiteness of \s$\Phi\m$.
\m Note that Eq.\s\s(\ref{ane}) may be rewritten 
by integration by parts and the Feynman-Kac formula (\ref{Brbr})
as
\qq
\Phi_t(\chi)\ =\s[\m\ee^{-t\m(-D'\Delta
+V_\chi)}\m 1\m](e)\s=\s
E_e\left(\ee^{-\int_0^t V_\chi(h(s))\s\m ds}\right)
\label{nnn}
\qqq
which follows also directly from Eq.\s\s(\ref{nnm}) if we
notice that the diffusion process \s$s\mapsto g_{t,t-s}\s$
on \s$SL(d)\s$ projects to the Brownian motion on \s$H_d\m$.
\m The resulting alternative expressions for \s$\Phi\s$:
\qq
\Phi(\chi)\ =\ \lim\limits_{t\to\infty}
\ \int\ee^{-t\m(-D'\Delta
+V_\chi)}(e,h)\s\m dh
\ =\ E_e\left(\ee^{-\int_0^\infty V_\chi(h(s))\s\m ds}\right).
\label{quad1}
\qqq
relate \s$\Phi(\chi)\s$ to the long time
behavior of the diffusion on the homogeneous space \s$H_d\s$ 
in the presence of a positive potential \s$V_\chi\s$ 
or to the low-energy properties 
of the Schr\"{o}dinger operator \s$-D'\Delta\m+\m V_\chi\m$.
\m They imply that \s$0\leq\Phi(\chi)\leq 1\m$. \m Expressions
(\ref{quad1}) may also be obtained directly in the 
Martin-Siggia-Rose (MSR) \cite{MSR} formal functional integral 
approach.
\vskip 0.5cm

By Minlos Theorem, the normalized (\m$\Phi(0)=1\m$),
continuous, positive-definite functional \s$\Phi\s$ 
on \s$\CS(\NR^d)\s$ given by
Eqs.\s\s(\ref{quad}) or (\ref{quad1}) defines a unique
probability measure \s$d\mu\s$ on  \s$\CS'(\NR^d)\s$ s.\m t.
\qq
\Phi(\chi)\ =\ 
\int\ee^{\m i\int T(x)\s\chi(x)\s dx}\s d\mu(T)\s.
\label{Minl}
\qqq
$d\mu$ is the stationary state of the Kraichnan model for $\gamma=0$
alluded to in Sect.\s\s7. 
It is quite different from the Gibbs measure and quite non-Gaussian
and is, indeed, supported by distributional configurations 
of the scalar since the correlation functions 
\qq
\CF_{2n}(\Nx)\ =\ \int T(x_1)\s\cdots\s T(x_{2n})\s\s d\mu(T)
\qqq
diverge logarithmically at coinciding points. 
The measure \s$d\mu\s$ contains all the joint p.d.f.'s 
of smeared scalar values \s$\int T(x)\s\chi(x)\s dx\m$.
In particular, the function \s$p\mapsto\Phi(p\chi)\m$,
is the Fourier transform of the p.d.f. $p_\chi(\theta)\s$ 
of \s$\int T(x)\s\chi(x)\s dx\s$ whose behavior was studied
in \cite{FKLM}, see also \cite{SS1}\cite{CFKL}\cite{CGK}.
\vskip 0.5cm

\s$\Phi(p\chi)\s$ is a pointwise 
limit of the finite-time functions \s$\Phi_t(p\chi)\s$
which are entire in \s$p\m$. \m For \s${\rm Re}\s p^2\geq -b^2\s$, 
\s$b>0\m$, 
\qq
\vert\m\Phi_t(p\chi)\m\vert\s\leq\s\Phi_t(\pm ib\chi)
\s=\s E_e\left(\ee^{\s b^2\int_0^tV_\chi(h(s))\s\m ds}
\right)\s=\s\int\ee^{\m t\m (D'\Delta\m+\m b^2 
V_\chi)}(e,h)\s\s dh\s\m\cr
=\ 1\s+\s b^2\int_{_0}^{^t}
\hs{-0.2cm}ds\int\ee^{\m s\m(D'\Delta
\m+\m b^2V_\chi)}(e,h)\s\m V_\chi(h)\s\s dh\s.
\label{toss}
\qqq
The Schr\"odinger operator \s$-D'\Delta\m-\m b^2V_\chi\s$ with 
a negative potential may develop bound states.
The right hand side of the inequality (\ref{toss}) grows with 
\s$t\s$ since the expression under the integrals is positive. 
If \s$e_b\equiv{\rm inf}\{{\rm spec}(-D'\Delta\m-
\m b^2V_\chi)\}<0\s$ then the growth 
is unbounded since \s$\ee^{\m s\m (D'\Delta\m+\m b^2 V_\chi)}(e,h)
\s\sim\s\ee^{-s\s e_b}\s$ for large \s$s\m$. \m On the other hand,
for \s$e_b>0\s$ the right hand side of (\ref{toss})
would be bounded uniformly in \s$t\s$ if \s$V_\chi\s$
were of compact support on \s$H_d\m$. \s$V_\chi\m$,  
\m however, does not have a compact support as a function
on \s$H_d\s$ even if \s$\CC\s$ and \s$\chi\m$ do (if they 
do not vanish identically). It is, nevertheless, easy to see 
from the definition (\ref{Vchi}) that \s$V_\chi\s$ vanishes
at infinity of \s$H_d\m$, i.e.\s\s\m that it gets arbitrarily 
small outside sufficiently big compact subsets of \s$H_d\m$.
\m This is enough to assure a uniform bound for the right hand 
side of (\ref{toss}) as may be seen by the
following argument which separates the behavior at infinity 
of \s$H_d\s$ from that in the interior. 
Write \s$V_\chi=V_\chi'+V_\chi''\s$ 
where $0\leq V_\chi'\leq V_\chi\s$ and \s$V_\chi'$ 
has a compact support. By 
the H\"{o}lder inequality, 
\qq
E_e\left(\ee^{\s b^2\int_0^tV_\chi(h(s))\s\m ds}
\right)\s\leq\s
E_e\left(\ee^{\s(1+\epsilon)\m b^2\int_0^tV_\chi'(h(s))\s\m ds}
\right)^{_1\over^{1+\epsilon}}
E_e\left(\ee^{\s{1+\epsilon\over\epsilon}\m 
b^2\int_0^tV_\chi''(h(s))\s\m ds}
\right)^{_\epsilon\over^{1+\epsilon}}.\hspace{0.7cm}
\label{Hoel}
\qqq
If we choose \s$\epsilon\s$ small so 
that for \s$b'=(1+\epsilon)^{\hf}b\s$ 
the relation \s$e_{b'}>0\s$ still holds then the first expectation
on the right hand side of inequality (\ref{Hoel}) is bounded
uniformly in \s$t\m$ (\m$e_b\s$ increases with decrease 
of \s$V_\chi\m$). \m Choose the support of \s$V_\chi'\s$
so that \s${1+\epsilon\over\epsilon}\m 
b^2\m V_\chi''\leq v_0<{Dd^2\over 4}\s$ where \s$v_0\s$
is a constant. Then
\qq
&&E_e\left(\ee^{\s{1+\epsilon\over\epsilon}\m 
b^2\int_0^tV_\chi''(h(s))\s\m ds}
\right)\s\leq\ 1\s+\s{_{1+\epsilon}\over^\epsilon}\m 
b^2\int_{_0}^{^t}\hs{-0.2cm}ds\int\ee^{-s\m(D'\Delta-v_0)}(e,h)
\s\s V_\chi(h)\s\s dh\cr
&&=\ 1\s+\s{_{1+\epsilon}\over^{2\m\epsilon}}\m 
b^2\int_{_0}^{^t}\hs{-0.2cm}ds\int
\ee^{-s\m(M_2-v_0)}
(x_{12}-y)\s\s\CC(y)\s\chi(x_1)\s\chi(x_2)\s\s dx_1\s dx_2\s dy
\qqq
and the last expression is bounded uniformly in \s$t\s$
as may be easily seen from Eq.\s\s(\ref{aa}).
\vskip 0.4cm

We infer that \s$\Phi(p\chi)\m$, as a limit 
of uniformly bounded analytic functions, is analytic
in \s$p\s$ for \s${\rm Re}\s p^2>b_0^2\s$ but has a singularity
at \s$p=\pm ib_0\s$ where \s$b_0\s$ is the positive number 
s.t. \s$e_{b_0}=0\m$. \m The Cauchy bounds imply now that 
\qq
|\m{_{d^n}\over^{dp^n}}\s\Phi(p\chi)\m|\ \leq\ 
{_{{\rm const}.(\epsilon)\s n!}\over^{1+|p|^n}}
\qqq
in any strip \s$|{\rm Im}\s p|\m<\m b_0-\epsilon\s$
for \s$\epsilon>0\m$. \m Note also that 
\s$\lim\limits_{\vert p\vert\to\infty}\ \Phi(p\chi)\ =\ 0\s$
by virtue of Eq.\s\s(\ref{quad1}). Since
\qq
p_\chi(\theta)\s=\s{_1\over^{2\pi}}\int\ee^{-i\theta p}\s\m
\Phi(p\chi)\s\s dp\s,
\label{FTr}
\qqq
it is easy to show integrating by parts few times
and moving the \s$p$-integration contour to \s${\rm Im}\s p=
\pm(b_0-\epsilon)\s$ that \s$p_\chi(\theta)\s$ is smooth
except, possibly, at \s$\theta=0\s$ and that
\qq
p_\chi(\theta)\ \leq\ {\rm const.(\epsilon)}\ \ee^{-(b_0-\epsilon)
\vert\theta\vert}
\qqq
for \s$|\theta|\geq\CO(1)\s$ and any positive \s$\epsilon\m$.  
\m Clearly, the same inequality fails 
for negative \s$\epsilon\s$ since it would imply analyticity of
\s$\Phi(p\chi)\s$ at \s$p=\pm ib_0\m$. \s {\bf In short}: the p.d.f. 
\m$p_\chi(\theta)\s$ of \s$\int T(x)\s\chi(x)\s dx\s$ has an
exponential decay for large \s$\vert\theta\vert\s$ 
with the rate \s$b_0\s$ equal to the value of \s$b\s$ at which 
the ground state of \s$-D'\Delta-b^2V_\chi\s$ crosses zero energy.
Note that the rate \s$b_0\m$, as related to a bound state energy is
not, in general, a semi-classical quantity. 
\vskip 0.4cm

For rotationally invariant \s$\chi\m$ which, for simplicity,
we shall normalize so that \s$\int\chi=1\m$, \m our operators
on \s$H_d\s$ reduce to the ones on the double coset space
\s$SO(d)\backslash SL(d)/SO(d)\m$. \m This space
may be identified with the Cartan algebra of \s$SL(d)\s$ 
divided by the action of the Weyl group and may be parametrized
by the diagonal matrices \s${\rm diag}(\phi_1,\dots,\phi_d)\s$ 
with entries \s$\phi_1\leq\cdots\leq\phi_d\s$
and s.t. \m$\sum_i\phi_i=0\m.$ 
\m In this parametrization, the Schr\"{o}dinger 
operator \s$-D'\Delta+b^2V_\chi\s$ becomes the Calogero-Sutherland 
Hamiltonian \cite{CS}\cite{PO} with a potential:                               
\qq
D'\left(-\m\hf\sum\limits_i{_{d^2}\over^{d\phi_i^2}}\s-
\s\sum\limits_{i<j}{_1\over^{4\m\sinh^2(\phi_j-\phi_i)}}\s+
\s{_{d(d^2-1)}\over^{24}}\right)-\s b^2\m V_\chi(\phi_i)\s,
\label{SL}
\qqq
acting in \s$L^2(\prod\limits_i^{d-1} d\phi_i)\m$. \s The constant 
\s${d(d^2-1)\over24}\m$, \m equal to the half length squared 
of the Weyl vector of \m$SL(d)\m$, \m is the infimum 
of the spectrum of \s$-\Delta\s$ \cite{PO} so that 
\s$e_b\geq{D'd(d^2-1)\over 24}-b^2 V_\chi(0)\m$. 
\m Note that, for \s$d>2\m$, \s${d(d^2-1)\over24}\s$
is higher than the infimum of the spectrum of \s$-\hf H^2\s$
acting in \s$L^2(R^d)\s$ since, as pointed out in the remark 
after Eq.\s\s(\ref{tbcl}), the latter is equal to \s${d(d-1)}
\over 8\m$. \m This discrepancy 
is due to the appearence of different irreducible 
representations in the decomposition
of the actions of \s$SL(d)\s$ in \s$L^2(H_d)\s$ and in 
\s$L^2(\NR^d)\s$ for \s$d>2\s$ and it will play an important 
role below.
%The two-dimensional 
%case is specially simple. We obtain a differential operator 
%in the single variable 
%\s$\phi\equiv \phi_2-\phi_1\m$, \s$\phi\geq 0\m$, 
%\m with the potential \s$-b^2V_\chi(\phi)\s$ regular at \s$\phi=0\s$ 
%and decaying as \s$\ee^{-\phi/2}\s$ at infinity. The Green function 
%\s$G_b(e,h)\s$ is represented by the solution \s$\tilde G_b(\phi)\s$ 
%of the equation
%\qq
%D'\left(-\m{_{d^2}\over^{dq^2}}\s-
%\s{_1\over^{4\m\sinh^2(\phi)}}\s+
%\s{_{1}\over^4}\right) \tilde G_b(\phi)\s-\s b^2\m 
%V_\chi(\phi)\s\s 
%\tilde G_b(\phi)
%\ =\ 0
%\qqq
%such that \s$\tilde G_b(\phi)\cong{_1\over^{D'}}\m 
%\phi^{1/2}\ln \phi\s$ 
%around \s$\phi=0\s$ and \s$\tilde G_b(\phi)\sim\ee^{-\phi/2}\s$ 
%as \s$\phi\to\infty\m$. \m Such a solution exists as long
%as \s$e_b>0\s$ by the standard Sturm-Liouville theory
%and the expression (\ref{drp}) which becomes 
%\qq
%1\s+\s b^2\int_{_0}^{^\infty}\hs{-0.15cm}\tilde G_b(\phi)\s\s 
%V_\chi(\phi)\s\s\sinh^{1/2}\phi\s\s d\phi
%\qqq
%is finite. For general \s$d\m$, \m the integral in (\ref{drp}) 
%is equal to
%\qq
%\int\limits_{\phi_1\leq\cdots\leq \phi_d}
%\hs{-0.2cm}\tilde G_b(q_i)\s\s 
%V_\chi(q_i)\s\s\prod\limits_{i<j}
%\sinh^{1/2}(q_j-q_i)\s\s
%\delta(\sum_{_i}q_i)\s\s\prod\limits_idq_i\s\s
%\qqq
%with \s$\tilde G_b(\phi_i)\s$ behaving as \s$\sim\m\prod\limits_{i<j}
%\ee^{\m\hf(\phi_i-\phi_j)}\s$ at infinity and bounded around 
%\s$\phi_i=\phi_j\s$ so that (\ref{drp}) is again finite as long as
%\s$e_b>0\m$.
%\vskip 0.4cm
\m When the forcing covariance 
\s$\CC(r)\s$ is essentially constant for 
\s$r\m\mathop{<}\limits_{^\sim}\m L\s$ and is falling off to zero 
for \s$r\m\gg\m L\s$ (e.g. for \s$\CC(r)\s$ replaced by
\s$\CC_L(r)\equiv\CC(r/L)\m$) 
\m then the potential \s$-b^2V_\chi\s$
approaches for \s$L\to\infty\s$ a constant equal to 
\s$-\hf b^2\CC(0)\s$ so that for large \s$L\s$ we obtain
\s$\hf\s b_0^2\s \CC(0)\s\cong\s{_{D'd(d^2-1)}\over^{24}}\s$ or 
\qq
b_0\m\cong\m d\s\sqrt{_{D(d+1)}\over^{6\m\CC(0)}}\s.
\qqq
Note however that although \s$b_0\s$ stabilizes when \s$L\to\infty\m$,
\m the right hand side of Eq.\s\s(\ref{toss}) tends to 
\s$\ee^{\m\hf\m t\m b^2\m \CC(0)}\s$ and blows up with \s$t\m$. 
\vskip 0.4cm

The exponential decay of the scalar p.d.f. for \s$\gamma=0\s$ 
in the isotropic two-dimensional situation was first found in \cite{CFKL},
see also \cite{SS1} for a discussion of the non-isotropic case. 
The calculation of \cite{CFKL} was extended to higher dimensions
in \cite{CGK}. Both calculations were reinterpreted in \cite{FKLM}
within the semiclassical approach. Our rigorous result about 
the decay rate \s$b_0\s$ of \s$p_\chi(\theta)\s$ disagrees for 
\s$d>2\s$ with the result of \cite{CGK} and with the instanton 
calculation of \cite{FKLM}. These papers obtain the value 
\s$b_0'=d\s\sqrt{D\over2\m\CC(0)}\m$ for the decay rate 
which is smaller than \s$b_0\s$ for \s$d>2\m$. 
\m The point is that in \cite{CFKL} and \cite{CGK} the function 
\s$V_\chi(g)\s$ of Eq.\s\s(\ref{Vchi}) was replaced by 
\s$V_{x}(g)=\hf\s\CC(g^{-1}x)\s$ with fixed \s$x\not=0\m$. 
\m This simplifies the calculation of the expressions of 
Eq.\s\s(\ref{quad1}) since only the distribution of \s$g_{t,s}x\s$ 
for one \s$x\s$ is needed. They become
\qq
\Phi'_x\ \equiv\ 
E_e\left(\ee^{-\int_0^\infty V_x(h(s))\s\m ds}\right)
\ =\ \lim\limits_{t\to\infty}
\ \int_{_{\NR^d}}\ee^{-t\m(M_2\m+\m\hf\m\CC)}(x,y)\s\m dy
\label{quad2}
\qqq
and lead to the quantum mechanical problem analyzed in
\cite{CFKL}\cite{CGK}. Upon the replacement 
of \s$V_x\s$ by \s$p^2\s V_x\s$ one obtains
a function \s$\Phi'_x(p)\s$ whose first singularity off the real 
axis is at \s$p=\pm ib\s$ with \s$b\s$ s.\m t. the ground state 
of \s$M_2-\hf\m b^2\CC\s$ crosses zero energy. Since
the spectrum of \s$M_2\s$ starts from \s${D\m d^2\over 4}\m$,
\m see the remark after Eq.\s\s (\ref{tbcl}), 
we indeed obtain, for \s$\CC=\CC_L\s$ and large \s$L\m$, 
\m the exponential decay rate \s$b'_0\s$ for the Fourier transform
of \s$\Phi'_x(p)\m$. \m The technical reason for the discrepancy 
with our exact calculation is that \s$V_x(g)\m$, \m unlike its
smeared version \s$V_\chi(g)\s$, does not vanish at the infinity 
of \s$H_d\m$ and leads to a more singular behavior of the right 
hand side of Eq.\s\s(\ref{toss}). Another way to see it\footnote{we 
thank M. Chertkov for suggesting this interpretation} is that 
\s$\Phi_x'\s$ is given by a version of Eq.\s\s(\ref{quad0})
with \s$\int\prod\chi(x_i)\s dx_i\s$ omitted and with
\s$F_{2n}(\Nu_p)\s$ replaced by the partition-independent
contribution \s$F_{2n}(x,\dots,x)\s$ corresponding
the collinear configuration \s$\Nu_p\s$ giving the most singular
behavior when \s$\Nu_p\to0\s$ (see Appendix B). The 
smearing in Eq.\s\s(\ref{quad0}) makes this behavior more 
regular. Our result persists, however, also if we replace
\s$V_\chi(g)\s$ with \s$\tilde V_\psi(g)
=\hf\int\CC(g^{-1}x)\s\psi(x)\s dx\m$,
\m if $\psi$ and \s$\CC\s$ are non-negative function from 
\s$\CS(\NR^d)\m$, \m since \s$\tilde V_\psi(g)\s$ still vanishes 
at infinity. In particular, \s$\psi\s$ may vanish around the origin 
which shows that it is the smearing of collinearity, not the inclusion 
of coinciding points, which is responsible for the discrepancy between 
\s$b_0\s$ and \s$b_0'\m$. \m The lesson is that the correlation 
of (non-collinear pairs of) Lagrangian trajectories renders the smeared 
scalar less intermittent in more than two dimensions and should 
not be neglected.
\vskip 0.5cm

It is easy to see that \s$\Phi(p\chi)\s$
decays exponentially for large real \s$p\m$.
\m Denote by \s$\tau\s$ the first exit time
of the Brownian motion on \s$H_d\s$ from a fixed neighborhood
of \s$e\s$. The probability of a given value of \s$\tau\s$
is bounded by \s$\ee^{-{\rm const}./\tau}\m$.
\m Since \s$V\equiv\int_0^\infty\hs{-0.1cm}V_\chi(h(s))\s ds 
\m\geq\m{\rm const}.\s\tau\m$, \m the conditional expectation 
\s$E_e(\ee^{-p^2V}\m\vert\m\tau)\s$ is bounded 
by \s$\ee^{-{\rm const}.\m p^2\tau}\m$. 
\m Hence the exponential decay of \s$E_e(\ee^{-p^2 V})
\leq\int_0^\infty\hs{-0.1cm}\ee^{-{\rm const}.\m
(p^2\tau\m+\m1/\tau)}\s\m d\tau\m$. \s A more exact 
description of the decay follows from the path-integral 
integral representation of the expectation (\ref{quad1}). 
The latter implies that the large \s$p\s$ behavior 
of \s$\Phi(p\chi)\s$ for real \s$p\m$, \m unlike the large 
\s$\theta\s$ behavior of \s$p_\chi(\theta)\m$, 
is semi-classical:  
\qq
\Phi(p\chi)\ \sim\ \ee^{-\m |p|\s S(g(\m\cdot\m))}
\qqq
where \s$[0,\infty]\mapsto g(s)\s$ describes a trajectory 
(instanton) in \s$H_d\s$ minimizing the action 
\qq
S(h(\m\cdot\m))\s=\s\int_{_0}^{^\infty}
\hs{-0.21cm}\left({_1\over^{2D'}}\m
|\s\s{^{^{\cdot}}}\hs{-0.22cm}h(s)|^2\s+\s V_\chi(h(s))\right)ds
\qqq
\m for fixed initial 
value \m$h(0)=e\s$  (with \s$|\m\cdot\m|^2\s$
standing for the $SL(d)$-invariant metric on \s$H_d\m$). 
\m This is the same instanton as in the field theoretic 
MSR analysis of \cite{FKLM}.
For rotationally invariant \s$\chi\m$, the problem reduces to 
the one on \s$SO(d)\backslash SL(d)/SO(d)\s$
with the action
\qq
S(\phi_i(\m\cdot\m))\s=\s\int_{_0}^{^\infty}\hs{-0.2cm}
\left({_1\over^{2D'}}\m
\sum\s{^{^{\cdot}}}\hs{-0.18cm}\phi_i(s)^2\s+\s V_\chi(\phi_i(s))
\right)ds
\qqq
and with the initial value \s$\phi_i=0\m$. \m In \s$d=2\s$ 
the minimal value of \s$S\s$ is
\qq
{_1\over^{\sqrt{D'}}}\int_{_{0}}^{^\infty}\hs{-0.2cm}
\sqrt{V_\chi(\phi)}\s\s d\phi\ >\ 0
\qqq
where \s$\phi\equiv \phi_2-\phi_1\m$.
\m For \s$\CC(r)\s$ approximately constant up to \s$r\cong L\m$,
\s$V_\chi(\phi)\s$ is approximately constant up to \s$\hf\phi\cong
\ln{L}\s$ and then it decays to zero like \s$\sim\ee^{-\phi/2}\m$.
\m Consequently, the exponential decay rate of \s$\Phi(p\chi)\s$
is approximately \s$\ln{L}\s\s\sqrt{\CC(0)\over 2D}
\m$ for large \s$L\m$, \m in agreement with
\cite{CFKL} and \cite{FKLM}. For \s$d>2\m$ and large \s$L\s$ the minimum
of \s$S\s$ is attained on the trajectory which in the region
of constant potential goes in the direction 
\s$\sqrt{1\over^{d(d-1)}}(-1,\dots,-1,d-1)\s$ and the value of  
the action is again \s$\cong\m\ln{L}\s\s\sqrt{\CC(0)\over 2D}
\m$ up to lower order terms, as pointed out
in \cite{CGK} and \cite{FKLM}. The exponential 
decay of \s$\Phi(p\chi)\s$ implies that \s$p_\chi(\theta)\s$ 
is smooth also at zero.
\vskip 0.8cm

\nsection{Conclusions}

In this paper we have analyzed the stochastic dynamics of 
Lagrangian trajectories for Gaussian, time-decorrelated 
random velocity fields considered in the Kraichnan model 
of passive advection. We found that the dynamics 
is characterized by two related phenomena. First, 
the Lagrangian trajectories loose in the limit of high 
Reynolds numbers the deterministic sense for a fixed velocity 
realization due to their sensitive dependence 
on initial conditions. Second, their relative stochastic
dynamics is dominated by slow resonance-type modes. 
The slow modes determine the average characteristics
of the spread of Lagrangian trajectories responsible for
the loss of their deterministic character.
Both phenomena were essentially due to non-smoothness
of the typical velocities signaled by fractional H\"{o}lder
exponents in their spatial dependence. Since
the turbulent velocities are non-smooth
in the limit of high Reynolds numbers, we expect the
two phenomena to persist for more realistic velocity
ensembles and to continue to be responsible for the anomalous
scaling. For the spatially smooth velocities, we 
calculated the Lyapunov exponents describing 
the sensitive dependence of the Lagrangian trajectories
on initial conditions for distances smaller than the viscous
scale.  Using harmonic analysis on the symmetric 
spaces \s$SL(d)/SO(d)\s$ we also obtained 
in this case an explicit form of the characteristic 
functional of the stationary state of the passive scalar
and exhibited an exponential decay of the p.d.f.'s 
of smeared values of the scalar relating the decay rate 
to the properties of the ground state of the Calogero-Sutherland
Schr\"{o}dinger operator with a potential.
\vskip 0.5cm
%\eject

\nappendix{A}
\vskip 0.3cm

\noindent We shall make explicit the structural result
of Sect.\s\s6 for the heat kernel \s$\ee^{-t\m M_2}(x,x_0)\m$.
In the angular momentum \s$l=0,1,\dots\s$ sector,
\qq
M_2\s\equiv\s M_2(l)\ =\ -\m{_D\over^{r^{d-1}}}\s\da_r\m
r^{d+1-\gamma}\da_r\s+\s{_{D(d+1-\gamma)}\over^{d-1}}\m l\m(d-2+l)\s 
r^{-\gamma}
\qqq
which is a positive operator in \s$L^2(]0,\infty[,\m r^{d-1}dr)\m$.
\m The generalized eigen-function of \s$M_2(l)\s$ 
corresponding to eigenvalue \s$E\geq0\s$ 
involves the Bessel function
\qq
\varphi_E(r)\ =\ r^{^{\gamma-d\over2}}\s J_{\nu_l}(
{_{2\sqrt{E/D}}\over^{\gamma}}\s r^{^{\gamma\over2}})
\qqq
where 
\qq
\nu_l\s=\s{_1\over^{\gamma}}\sqrt{(d-\gamma)^2\s+\s4\m{_{d+1-\gamma}
\over^{d-1}}\m l\m(d-2+l)}\s\m.
\qqq
The spectral decomposition of \s$M_2(l)\s$ has the form
\qq
M_2(l)\ =\ \int E\s\s|\varphi_E\rangle\langle\varphi_E|\s\s d\nu(E)\s.
\qqq
Since 
\qq
(\CU_s\varphi_E)(r)\s\equiv\s\ee^{\m sd/2}\m\varphi_E(\ee^s r)\ =\ 
\ee^{^{\m{_\gamma\over^2}\m s}}\s\varphi_{\ee^{\m\gamma\m s}E}(r)\s,
\qqq
we infer that
\qq
\CU_sM_2(l)\m\CU_s^{-1}\ =\ 
\int E\s\s|\varphi_E\rangle\langle\varphi_E|\s\s 
d\nu(\ee^{\m\gamma\m s}E)\s.
\qqq
Since, on the other hand, \s$\CU_sM_2\CU_s^{-1}\s=\s
\ee^{\m\gamma\m s}M_2\m$, \m see Eq.\s\s(\ref{crin}), it follows that
\qq
d\nu(\ee^{\m\gamma\m s}E)\s=\s\ee^{\m\gamma\m s} d\nu(E)\s,
\qqq
i.e.\s\s that \s$d\nu(E)=c\s dE\s$ for some positive constant \s$c\m$.
Hence
\qq
c\int|\varphi_E\rangle\langle\varphi_E|\s\s dE\ =\ I
\qqq
and for
\qq
\hat f(E)\ =\ \sqrt{c}\int_{_0}^{^\infty}
\overline{\varphi_E(r)}\s\m f(r)\s\m r^{d-1}\s dr\s,
\qqq
we obtain \s$\int_{_0}^{^\infty}|\hat f(E)|^2\s dE\s=\s
\int_{_0}^{^\infty}|f(r)|^2\s r^{d-1}\s dr\m$.
Substituting \s$E=\ee^{\m\gamma u}\m$, we shall define
\qq
(\CV_1 f)(u)\ =\ \sqrt{\gamma}\s\s\ee^{^{\m{{\gamma\over 2}}\m u}}\m
\hat f(\ee^{\m\gamma u})\s.
\qqq
\s$\CV_1:\s L^2(]0,\infty[,\m r^{d-1}dr)\s\rightarrow
\s L^2(\NR, du)\s$ is a unitary operator. Besides,
\qq
(\CV_1 M_2(l)\m f)(u)\ =\ \ee^{\m\gamma u}\m(\CV_1 f)(u)\s.
\label{eq11}
\qqq
Let \s$\CV_2:\s L^2(]0,\infty[,\m r^{d-1}dr)\s\rightarrow
\s L^2(\NR, du)\s$ be another unitary  operator defined by
\qq
(\CV_2 f)(u)\ =\ \ee^{^{-{d\over 2}u}}\m f(\ee^{-u})\s.
\qqq
Note that
\qq
(\CV_i\CU_s f)(u)\ =\ (\CV_i f)(u-s)\s,\quad i=1,2\s,
\qqq 
so that \s$U_2\s=\s\CV_1^{-1}\CV_2\s$ commutes with \s$\CU_s\m$.
\s Besides, 
\qq
M_2(l)\ =\ U_2\s r^{-\gamma}\m U_2^{-1}\s,
\qqq
as follows from Eq.\s\s(\ref{eq11}). This is the relation (\ref{struc})
for \s$n=2\m$. \m Since, explicitly,
\qq
(\CV_1f)(u)\ =\ \sqrt{\gamma c}\int\ee^{^{\m{\gamma\over2}(u-u')}}
\s\s J_{\nu_l}({_2\over^{\gamma\sqrt{D}}}\s\ee^{^{\m{\gamma\over2}
(u-u')}})\s\s(\CV_2f)(u')\s\s du'
\qqq
and the Mellin transform is the composition of \s$\CV_2\s$ and
the Fourier transform, we obtain for \s${\rm Re}\s\sigma={d\over2}\m$
\qq
&&\hat{U}_2(\sigma)^{-1}\ =\ \sqrt{\gamma c}\int
\ee^{^{({d\over 2}+\sigma)u}}\ \ee^{^{\m{\gamma\over2}u}}
\s\s J_{\nu_l}({_2\over^{\gamma\sqrt{D}}}\s\ee^{^{\m{\gamma\over2}
u}})\s\s du\ =\ \sqrt{\gamma cD}
\s\left({_\gamma\over^2}\sqrt{D}\right)^{^{d+2\sigma\over\gamma}}\cr
&&\hs{1cm}\cdot\ 
\int x^{^{d+2\sigma\over\gamma}}\s\m J_{\nu_l}(x)\s\s dx
\ =\ \sqrt{\gamma cD}\s\left(\gamma
\sqrt{D}\right)^{^{d+2\sigma\over\gamma}}
\s{\Gamma({1\over 2}(1+\nu_l+{d+2\sigma\over\gamma}))\over
\Gamma({1\over 2}(1+\nu_l-{d+2\sigma\over\gamma}))}\s.\hs{0.4cm}
\qqq
The unitarity implies that \s$\gamma cD=1\s$ so that, finally, 
\qq
\hat{U}_2(\sigma)\ =\ 
\left(\gamma\sqrt{D}\right)^{^{-{d+2\sigma\over\gamma}}}
\s\s{\Gamma({1\over 2}(1+\nu_l-{d+2\sigma\over\gamma}))\over
\Gamma({1\over 2}(1+\nu_l+{d+2\sigma\over\gamma}))}\s.
\qqq
The right hand side has a meromorphic continuation to the complex plane 
of \s$\sigma\s$ with poles at 
\qq
\sigma_{l,p}\ =\ -\m {_{d-\gamma}\over^2}\s+\s{_\gamma\over^2}\m\nu_l
\s+\s\gamma\m p
\qqq
for \s$p=0,1,\dots$ \m Since the true 
(more regular at the origin) zero mode of \s$M_2(l)\s$ 
occurs at scaling dimension \s$\sigma_{l,0}\m$, \m this 
is exactly the analytic structure predicted for 
\s$\hat{U}_n(\sigma)\m$. The function \s$r^{\sigma_l,p}\s$ 
(multiplied by an angular term) represents  
a slow 2-point mode in the angular momentum 
\s$l\s$ sector.
\vskip 0.5cm
%\eject

\nappendix{B}
\vskip 0.3cm

Let us briefly consider the convergence properties of
the integrals (\ref{F2n2}). Let \s$k_i\in SO(2)\s$ be
rotation matrices s.\m t. $u_i=k_i(r_i,0)\s$ where \s$r_i=|u_i|\m$.
\m We have $g(z_i)^{-1}k_i= (k_i^{-1}g(z_i))^{-1}=k'_i\m 
g(k_i^{-1}z_i)^{-1}$
for some \s$k'_i\in SO(2)\m$. \m Denoting
\s$\CC(k(r,0))\equiv\CC(r)\m$, \m observing that 
\s$|g(z_i)^{-1}(r_i,0)|=r_i y_i^{-{1\over 2}}\s$ and using
the \s$SL(2)\s$ invariance of $d\nu(z_i)\m$, \m we obtain
\qq
{}F_{2n}(\Nu)=\int G(i,z_1)\s G(\kappa_1z_1,z_2)\s\dots 
\s G(\kappa_{n-1}z_{n-1},z_n)\s
\prod\limits_i\CC({{r_i y_i^{-{1\over 2}}}})\s\s d\nu(z_i)
\label{A1}
\qqq
where \s$\kappa_i=k_{i+1}^{-1}k_{i}\s$ 
is the rotation by the angle between
\s$u_{i+1}\s$ and \s$u_{i}\m$.
We shall study the behavior of \s$F_{2n}\s$ as \s$r_i\s$ tend to 0. 
The following is a useful relation:
\qq
\int G(z,z')\s\m dx'={_1\over^{4D}}\s 
(\m y\s\theta(y'-y)\s+\s y'\s\theta(y-y')\m)
\s\equiv\s G_0(y,y')\s.
\label{G0}
\qqq
For the 4-point function, noting that \s${\rm Im}(\kappa_1 z_1)
=\gamma_1 y_1\s$ where
\qq
\gamma_1= [(x_1 \sin \vartheta 
+\cos\vartheta)^2+y_1^2\sin^2\vartheta]^{-1}\s, 
\qqq
$\vartheta$ being the angle between \s$u_2\s$ and \s$u_1\m$, 
\m we obtain
\qq
{}F_4(u_1,u_2)=\int G(i,z_1)\s G_0(\gamma_1 y_1,y_2)\s 
\CC({{r_1 y_1^{-{1\over 2}}}})\s\CC({{r_2 y_2^{-{1\over 2}}}})\s\s
y_2^{-2}dy_2\s\m d\nu(z_1)\s.
\qqq
Consider first the case \s$\vartheta=0\s$ i.e. $\gamma_1=1\m$. \m Then
\qq
{}F_4(u_1,u_2)&=&{_1\over^{(4D)^2}}\int (\theta(y_1-1)+y_1\m\theta(1-y_1))
(y_1\m\theta(y_2-y_1)+y_2\m\theta(y_1-y_2))\cr
&&\hspace{3cm}
\cdot\ \CC({{r_1 y_1^{-{1\over 2}}}})\s\CC({{r_2 y_2^{-{1\over 2}}}})\s
y_1^{-2}\m dy_1\m y_2^{-2}\m dy_2\s.
\qqq
Since $\CC=\CC_L\s$ has rapid decay at infinity, 
the integrals are effectively
cut to \s$y_i>(r_i/L)^2\s$ and produce logarithms of \s$(r_i/L)\s$ 
as these ratios tend to zero. The most singular contribution is 
from \s$y_1\m\theta(1-y_1)\s y_2\m\theta(y_1-y_2)\s$ term which 
yields \s$4\ln(r_1/L)\ln(r_2/L)-2(\ln(r_1/L))^2\s$
if \s$r_1>r_2\s$ and \s$2(\ln(r_2/L))^2\s$ if \s$r_2>r_1\m$. 
\m Thus 
\qq
{}F_{4,L}(u_1,u_2)+F_4(u_2,u_1)= ({_{\CC(0)}\over^{2D}})^2\m
\ln(r_1/L)\ln(r_2/L)\ +\ {\rm less \  singular}\label{lead}
\qqq
for \s$\vartheta =0\m$. 
\vskip 0.3cm

For \s$\vartheta\neq 0\m$, \m the \s$y_2\s$ integral yields
\qq
{}F_{4,L}(u_1,u_2)\ =\ {_1\over^{8\pi}}\s\int
\ln{x_1^2+(y_1+1)^2\over x_1^2+(y_1-1)^2}
\s \CC_L({{r_1 y_1^{-{1\over 2}}}})\s[\m(\ln (\gamma_1)\s+\s
\ln (y_1r_2^{-2}L^2))
\cr
\hspace{2cm}\cdot\ 
\vartheta(\gamma_1y_1-r_2^2/L^2)\s+\s B\m]\s\m y_1^{-2}\m dy_1\m dx_1
\qqq
where \s$B\s$ is bounded. For the \s$\ln (\gamma_1)\s$ term 
the only singularity
is at \s$y_1\s$ small and this term is bounded by \s${\rm const}.\m
|\ln(r_1/L)|\m$. \m The rest has same leading singularity 
as the \s$\vartheta=0\s$ calculation.
Thus recognizing in (\ref{lead}) the 2-point singularities 
(\ref{za1}), we infer that the leading singularity
of the 4-point function is Gaussian, the sum of products 
of 2-point functions.

The analysis of the general correlation is similar though tedious.
When all the points are on the same line i.e. all the angles
are zero, we can do all the \s$x_i\s$ integrals by (\ref{G0}). Most
singular contribution is the one where all \s$G_0(y_{i-1},y_i)\s$
are replaced by \s${1\over 4D}\m y_i\theta(y_{i-1}-y_i)\m$. \m Summing over
the permutations of the \s$u_i\s$ yields
\qq
\sum_\pi F_{2n}(\Nu_\pi)=\prod_i {_1\over^{4D}}\int_0^1
\CC_L(r_iy^{-\hf})\s y^{-1}dy\ +\ \dots
=\prod_i-\m{_{\CC(0)}\over^{2D}}\s\ln(r_i/L)\ 
+\ \dots\hspace{0.4cm}
\qqq
where \s$\dots\s$ is less singular. Non-zero angles give again
subleading contributions. The subsequent sum over unordered
pairings gives the Gaussian expression for the leading 
short-distance singularity of \s$\CF_{2n,L}(\Nx)\m$ in terms
of the singular contributions to the 2-point function,
in agreement with the observations of \cite{CFKL} and 
\cite{BCKL}.
\vspace{5mm}

Finally, for \s$d>2\s$, to extract the leading singularity
some bounds for \s$G\s$ are needed. Let us here note only
that for \s$d=3\s$ if all the \s$\kappa_i\s$ are identity, then
the \s$x\s$ integrals can again be done and the result
is that \s$G\s$ gets replaced by \s$G_0\s$ where
\qq
G_0 = {\rm const}.\s(-y_1^2\p_{y_1}^2-3y_2^2\p_{y_2}^2)^{-1}
\qqq
on \s$L^2((y_1 y_2)^{-2}dy_1dy_2)\s$ (we have put \s$y_1=\ee^\alpha\m$, 
\s$y_2=\ee^\beta\s$ in (\ref{lb})). The behavior of \s$G_0\s$
near \s$y_i=0\s$ is calculable and the leading singularity
can again be shown to be given by products of 2-point functions.
Using the Mellin transform we may write
\qq
(G_0F)(y'_1,y'_2)\s=\s
{\rm const}.\int_0^\infty\hs{-0.2cm}{_{dt}\over^t}
\s\ee^{-t}\int\hs{-0.1cm}{_{dy_1dy_2}
\over^{y_1y_2}}\s({_{y'_1y'_2}\over^{y_1y_2}})^{^{1\over 2}}
\s\m\ee^{-{1\over 4t}(\log{y_1
\over y'_1})^2 -{1\over 12t}(\log{y_2\over y'_2})^2}\s\m 
F(y_1,y_2)\s.\hspace{0.7cm}
\qqq
\s$\CC(ry^{-{1\over 2}})\s$ in (\ref{A1}) is replaced by 
\s$\CC(ry_1^{-{1\over 2}}
y_2^{-{1\over 6}})\m$. \m Hence, let $\rho = y_1^{{1\over 2}}
y_2^{{1\over 6}}$ and let $F(y_1,y_2)=f(\rho)$. Then
$(G_0F)(y'_1,y'_2)=\tilde{f}(\rho')$ with
\qq
\tilde{f}(\rho')\s=\s{\rm const}. 
\int_0^\infty {_{dt}\over^t}\s\m\ee^{-t}\int du\s dv
\s\s f(\ee^{-u-v}\rho')\s\m\ee^{u+3v}\s\m\ee^{-{u^2\over t}
-{3\m v^2\over t}}
\qqq
which after performing the \s$u-v\s$ 
and the \s$t\s$ integrals becomes
\s$\int g(\rho',\rho)\s f(\rho)\s d\rho\s$ with
\qq
g(\rho',\rho)={\rm const}.\s\m[\m
{_1\over^\rho}\s\theta(\rho'-\rho)+{_{{\rho'}^3}\over^{\rho^4}}
\s\theta(\rho-\rho')\m]\s.
\qqq
We may then write $F_{2n}$ in the form
\qq
{}F_{2n}(\Nu)\m=\int g(1,\rho_1)\m\dots\m g(\rho_{n-1},\rho_n)
\prod_{i=1}^n \CC(r_i\rho_i^{-1})d\rho_i
\qqq
and the analysis of the singularities goes on as in the
\s$d=2\s$ case.
%To show that \s$F_{2n}\s$ is well defined, let \s$C_x(g)=C(g^{-1}x)\m$. 
%\m Then
%\qq
%F_{2n}({\bf u})=(GC_{u_1}GC_{u_2}\dots GC_{u_n})(e).
%\qqq
%Using the \s$SO(3)$-invariance and setting \s$u'=(|u|,0,\dots ,0)\m$,
%\m we obtain
%\qq
%||GC_u||_{\infty}=||GC_{u'}||_{\infty}=||G_0\s C(y_1^{-{1\over 2}}
%y_2^{-{1\over 6}}|u|)||_{\infty}\cr
%=\sup_{\rho'}\int g(\rho',\rho)\s C(|u|\rho^{-1})\s\m d\rho
%\leq (A\m|\ln|u|\m|\m+\m B)\s.
%\qqq
%Hence
%\qq
%|F_{2n}({\bf u})|\s\leq\s\prod_{i=1}^n(A|\ln|u_i|\m|+B)\s.
%\qqq
%eject


\vskip 0.8cm

\begin{thebibliography}{bib}

\bibitem{Kolm}
A. N. Kolmogorov: {\it The Local Structure of Turbulence
in Incompressible Viscous Fluid for Very Large Reynolds'
Numbers}. C. R. Acad. Sci. URSS {\bf 30} (1941), 301-305

\bibitem{Kr68}
R. H. Kraichnan: {\it Small-Scale Structure of a Scalar
Field Convected by Turbulence}.
Phys. Fluids {\bf 11} (1968), 945-963

\bibitem{SS} 
B. Shraiman and E. Siggia:
{\it Anomalous Scaling of a Passive Scalar
in Turbulent Flow}. C.R. Acad. Sci. {\bf 321} (1995), 279-284

\bibitem{Falk}
M. Chertkov, G. Falkovich, I. Kolokolov and V. Lebedev: 
{\it Normal and Anomalous Scaling
of the Fourth-Order Correlation Function of a Randomly
Advected Scalar}. Phys. Rev. {\bf E 52} (1995), 4924-4941

\bibitem{GK}
K. Gaw\c{e}dzki and A. Kupiainen:
{\it Anomalous Scaling of the Passive Scalar}.
Phys. Rev. Lett. {\bf 75} (1995), 3834-3837

\bibitem{BGK}
D. Bernard, K. Gaw\c{e}dzki and A. Kupiainen:
{\it Anomalous Scaling in the N-Point Functions
of a Passive Scalar}. Phys. Rev. {\bf E 54} (1996), 2564-2572

\bibitem{SS1}
B. Shraiman and E. Siggia: {\it Lagrangian Path Integrals
and Fluctuations in Random Flow}. Phys. Rev. {\bf E 49} (1994),
2912-2927

\bibitem{SS2}
B. Shraiman and E. Siggia: {\it Symmetry and Scaling
of Turbulent Mixing}. Phys. Rev. Lett. {\bf 77} (1996), 2463-2466

\bibitem{CFKL}
M. Chertkov, G. Falkovich, I. Kolokolov and V. Lebedev: 
{\it Statistics of a Passive Scalar Advected by a Large-Scale
Two-Dimensional Velocity Field: Analytic Solution}.
Phys. Rev. {\bf E 51} (1995), 5609-5627

\bibitem{CGK}
M. Chertkov, A. Gamba and I. Kolokolov: {\it Exact Field-Theoretical
Description of Passive Scalar Convection in an $N$-Dimensional
Long-Range Velocity Field}. Phys. Lett. {\bf A 192} (1994), 435-443

\bibitem{GKo}
A. Gamba and I. Kolokolov: {\it The Lyapunov Spectrum of Continuous
Product of Random Matrices}. J. Stat. Phys. {\bf 85} (1996),
489-499

\bibitem{BCKL}
E. Balkovsky, M. Chertkov, I. Kolokolov and V. Lebedev:
{\it The Fourth-Order Correlation Function of a Randomly
Advected Passive Scalar}. Pis'ma v ZhETF, {\bf 61} (1995),
1012-1017

\bibitem{FKLM}
G. Falkovich, I. Kolokolov, V. Lebedev and A. Migdal:
{\it Instantons and Intermittency}. Phys. Rev. {E \bf 54}
(1996), 4896-4907

\bibitem{Chert}
M. Chertkov: {\it Instanton for Random Advection}. chao-dyn/9606011

\bibitem{FKL}
G. Falkovich, V. Kazakov and V. Lebedev. {\it Particle Dispertion
in a Multidimensional Random Flow with Arbitrary Temporal
Correlations}. To appear in J. Phys. {\bf A}

\bibitem{BeLv}
V. I. Belinicher, V. S. L'vov: {\it A Scale Invariant Theory
of Fully Developed Hydrodynamic Turbulence}. Sov. Phys. JETP
{\bf 66} (1987), 303-313

\bibitem{Falk2}
M. Chertkov and G. Falkovich: {\it Anomalous Scaling 
Exponents of a White-Advected Passive Scalar}. Phys. Rev. Lett {\bf 76}
(1996), 2706-2709

\bibitem{ca}
K. Gaw\c{e}dzki: {\it Turbulence under a Magnifying Glass}.
chao-dyn/9610003

\bibitem{Furst}
H. Furstenberg and H. Kesten: {\it Products of Random
Matrices}. Ann. Math. Statist. {\bf 31} (1960), 457-469

\bibitem{Past}
I. Ya. Gol'dshein, S. A. Molchanov and L. A. Pastur:
{\it A Pure Point Spectrum of the Stochastic One-Dimensional
Schr\"{o}dinger Operators}. Funct. Anal. Appl. {\bf 11} (1977), 1-8

\bibitem{Schlad}
K. Gaw\c{e}dzki and A. Kupiainen:
{\it Universality in Turbulence: an
Exactly Soluble Model}. In: Low-dimensional models
in statistical physics and quantum field theory.
H. Grosse and L. Pittner (eds.), Springer, 
Berlin 1996, pp. 71-105

\bibitem{SV}
D. W. Stroock and S. R. S. Varadhan: {\it Diffusion Processes
with Continuous Coefficients, I}. Commun. Pure Appl. Math.
{\bf 22} (1969), 345-400

\bibitem{MSR}
P. C. Martin, E. D. Siggia and H. A. Rose: {\it Statistical Dynamics
of Classical Systems}. Phys. Rev. {\bf A 8} (1973), 423-437

\bibitem{CS}
B. Sutherland: {\it Exact Results for a Quantum Many-Body Problem 
in One Dimension. II}. Phys. Rev. {\bf A 5} (1972), 1372-1376

\bibitem{PO}
M. A. Olshanetsky and A. M. Perelomov: {\it Quantum Integrable
Systems Related to Lie Algebras}. Phys. Rep. {\bf 94} (1983),
313-404

\end{thebibliography}
\end{document}